\documentclass[11pt,a4paper]{article}
\pdfoutput=1
\usepackage{jheppub}

\def\m{\mu}
\def\n{\nu}

\def\a{\alpha}
\def\b{\beta}
\def\s{\sigma}

\def\la{\lambda}
\def\La{\Lambda}

\def\Ups{\Upsilon}

\def\nd{{\rm{nd}}}

\def\GB{{\mathrm{GB}}}

\def\rm{\mathrm}
\def\cal{\mathcal}

\def\pa{\partial}

\def\be{\begin{equation}}
\def\ee{\end{equation}}
\def\br{\begin{eqnarray}}
\def\er{\end{eqnarray}}
\def\bsub{\begin{subequations}}
\def\esub{\end{subequations}}

\def\ch{\check}
\def\ha{\hat}
\def\til{\tilde}

\def\bul{$\;\bullet\;$}

\title{Scale Factor Duality  for  Conformal Cyclic Cosmologies } 
\author[a]{U. Camara dS,} 
\author[a]{A.L. Alves Lima}
\author[a]{and G.M. Sotkov} 
\affiliation[a]{Departamento de F\'\i sica - CCE\\
Universidade Federal de Espirito Santo\\
29075-900, Vitoria - ES, Brazil} 
\emailAdd{ulyssescamara@gmail.com} 
\emailAdd{andrealves.fis@gmail.com}
\emailAdd{gsotkov@yahoo.com.br}

\abstract{The scale factor duality  is  a symmetry of  dilaton gravity which is known to lead  to  pre-big-bang  cosmologies.  A  conformal time  version of the scale factor duality (SFD) was recently implemented   as a UV/IR symmetry between decelerated and accelerated phases of the post-big-bang evolution within  Einstein gravity coupled to a scalar field. The problem  investigated  in the present paper concerns 
the employment of  the conformal time SFD methods to the construction of  pre-big-bang and cyclic extensions of these models. We demonstrate that each big-bang model gives rise to two  qualitatively different  pre-big-bang evolutions: a  contraction/expansion SFD model and  Penrose's Conformal Cyclic Cosmology (CCC).  A few examples of SFD symmetric cyclic universes involving certain gauged K\"ahler sigma models  minimally coupled to Einstein gravity 
are studied. We also describe the  specific SFD features of the thermodynamics  and the conditions for  validity of the generalized second law in the case of Gauss-Bonnet (GB) extension of these selected  CCC models. }

\keywords{scale factor duality, pre-big-bang, conformal cyclic cosmology, Gauss-Bonnet gravity thermodynamics}

\begin{document} 
\maketitle 

\section{Introduction}\label{Sect.Intro}

Recent accumulation of observational evidences \cite{planck}  in favor of certain  inflationary improvements of the $\Lambda$CDM model has  stimulated  the search for the symmetry principles 
governing  the universe evolution \cite{thooft,hooft2015singularities,khoury1,kaloper,bars,linde2,lindeUniversality}.  An important advance  has been reached in the construction  of a few  selected  families of  inflaton  superpotentials  (representing  universal attractors \cite{linde1,LindeLargeFieldInflation, linde3}) within the frameworks of  (conformal)  supergravity \cite{ferrara, ferraraNMSSM}.  

The  theoretical interpretations of the new astrophysical data do not, however, completely rule out  the eventual physical relevance of appropriately chosen   ``pre-big-bang''  \cite{venez, VezianoPrebigbangincosmo}  or  ``cyclic'' \cite{turok, penrose, steinhardt2002cosmic, linde-cyclic-rev}  extensions of the standard $\Lambda$CDM universe. 
In these models the universe exists previously to the big-bang, in a phase of accelerated or/and decelerated contraction or  expansion.
The transition between standard  post-big-bang evolution and these pre-big-bang phases may be achieved by non-singular or singular bounces (leading to a big-crunch/big-bang ``joint''  \cite{turok-seiberg}), or else by the conformal identification of the future space-like infinity of the ``previous universe'' with our big-bang singularity,  as in  the \emph{conformal cyclic cosmology} (CCC) models \cite{penrose}.
The conceptual advantage of such models is that there is no beginning of time and the very special state seen at the big-bang  arises as a \emph{result} of the pre-big-bang evolution. The challenge consists in arranging  this previous evolution in order to address all  the problems that are usually solved by inflation, without introducing  any inconsistencies with the null energy condition (NEC) and with the  increasing of entropy, i.e.  with the generalized second law.  
 Within the vast variety of  cyclic cosmological models, characterized by the big-crunch-to-big-bang joints, one should also  stand out the cyclic extensions of the post-big-bang  eternal  (chaotic)  inflationary models with appropriate matter potentials, admitting regions of negative values \cite{linde-cyclic-rev, linde-kofman-bicyclic}.

Symmetry requirements of a ``stringy'' origin are known to be  an important tool in the construction of  pre-big-bang models \cite{venez, VezianoPrebigbangincosmo}. They allow one to define extensions of the Friedmann-Robertson-Walker (FRW) solutions before the  singularity, and to determine  the explicit form of the pre-big-bang matter content: e.g., they fix the equation of state (EoS)  of the fluid, or the potential of the corresponding inflaton or quintessence scalar fields before the big-bang. The most representative example  of such symmetries is given by the \emph{scale factor duality}  (SFD) ---  an extension of the superstring  T-duality \cite{giveon} for time dependent cosmological  backgrounds in dilaton gravity,  realized as a combination of scale factor inversions  together with time reflections, preserving the form of the Friedmann equations. It  provides  alternative pre-big-bang models that admit different cosmic and conformal time realizations, both in Einstein and in dilaton gravities \cite{venez,camara,dabrow,chim-wz, Chimento:2008qq}.


The present paper is devoted to the further investigation of the cosmological applications  of  a \emph{conformal time} scale factor duality \cite{camara},   
in FRW universes with metric
\br
ds^2=a^2(\eta) \left( -d\eta^2 + \frac{dr^2}{1-Kr^2} + r^2 d\theta^2 + r^2 \sin^2\theta \, d\varphi^2 \right) \label{frwmet}. 
\er
Here the spatial curvature is determined by $K = 0 , \pm 1$, we denote $a'=\frac{da}{d\eta}$, etc., in what follows, and  use units in which $8\pi G = 1$. 
 The duality transformations are composed of the scale-factor inversion
\bsub\label{sfd}\br
\til{a}(\til{\eta})=c_0^2/a(\eta), \quad\quad {\text{with}} \quad  \til{\eta}= \pm \eta + {\text{constant}} ,\label{sfinv}
\er
followed by the specific transformation for the  energy densities $\rho$ and $\til{\rho}$ and the pressures $p$ and $\til{p}$ of the matter fluids:
\br
  \til {a}^2 \til{\rho}(\til a)= a^2 \rho(a),\quad \quad \til {a}^2 [3\til {p}(\til a)+ \til{\rho}(\til a)] = -a^2 [3p(a)+\rho(a)] ,
 \label{sfdro}
\er
that may be combined into an expression for the equation of state parameters $w(\rho) \equiv p / \rho$ of the fluids, reading
\br
w(\rho)+\til w(\til{\rho})= -\tfrac{2}{3} . \label{sfdw}
\er
\esub
The duality acts in Einstein (and in Gauss-Bonnet) gravity, and is a symmetry of the space of solutions of the Friedmann equations
\br
  \frac{1}{3} \rho= \left(\frac{a'}{a^2}\right)^2+\frac{K}{a^2} \; , \quad
  p = \left(\frac{a'}{a^2}\right)^2 - \frac{2a''}{a^3} -\frac{K}{a^2} \; ,  \quad \rho'+3\frac{a'}{a}(\rho+p)=0 . \label{frw} 
\er
If one knows the evolution of a FRW universe described by a set $\{ a,  \rho,  w\}$, then one has automatically the solution of Eqs.(\ref{frw}) for the `dual'  universe described by $\{\tilde a, \tilde \rho, \tilde w\}$ as well.
As one can see directly from (\ref{sfdw}),  \emph{SFD maps accelerated periods into decelerated periods, and vice-versa} --- fluids such as radiation or dark matter, for which $w > - \frac{1}{3}$, are mapped into dark energy fluids with $w < - \tfrac{1}{3}$. 
 Further, \emph{the weak energy condition is preserved} for fluids with $-1 \leq w(\rho) \leq \frac{1}{3}$; that is, if $ w(\rho)$ belongs to this interval, then so will $\tilde w(\tilde\rho)$. On the other hand, the dual of a fluid with $w > \frac{1}{3}$ will be a phantom fluid with $\tilde w < -1$, so SFD also provides the possibility of describing phantom fluids from the knowledge of the behavior of non-phantom matter. 
 Recent  investigation \cite{camara} of the \emph{self-duality}  aspects of the  conformal time scale factor duality has demonstrated that it can play an important role in the description of the  \emph{post-big-bang} phases of the universe evolution, acting as a UV/IR symmetry relating the final (almost de Sitter) stage of the universe with its (radiation filled) beginning, and also determining the form of the matter self-interaction. 

 The problem we address  concerns the construction of pre-big-bang and cyclic  universes using  conformal time SFD (\ref{sfd}) as a symmetry principle. That is, given a model of post-big-bang cosmic evolution then, following Refs.\cite{venez, VezianoPrebigbangincosmo}, its pre-big-bang phase will be defined by the corresponding dual universe. This is equivalent to implement  the most relevant features of the pre-big-bang models of dilaton gravity, but here in  its Einstein-frame version
 and replacing the dilaton with an appropriate self-interacting scalar field,  
 as suggested in Ref.\cite{camara}. Consider as  an example the flat $\Lambda$CDM model and its dual:
\br
 \rho = \frac{ \rho_r}{ a^4} + \frac{ \rho_d}{ a^3} + \rho_{\Lambda}, \quad\quad   \tilde \rho =  \frac{\tilde \rho_r}{\tilde a^4} + \frac{\tilde \rho_{dw}}{\tilde a} + \tilde\rho_{\Lambda} , \label{LaCDMdaul}
\er
with $\rho_{\Lambda} = \La$ being a positive cosmological constant, and $\rho_r$ and $\rho_d$ representing radiation and (barionic plus dark) matter, respectively. The dual, pre-big-bang universe is filled with another cosmological constant, radiation and a gas of domain walls represented by $\tilde \rho_{dw}$,
with the relative densities fixed, according to the first of Eqs.(\ref{sfdro}), as
\br
 \rho_r / \tilde \rho_{\Lambda}  = \tilde \rho_r / \rho_{\Lambda} = \left( \rho_{d} / \tilde \rho_{dw} \right)^2 = c_0^4 . \label{densreladualLACDM}
\er
Our main goal is to describe how, although with the same dual matter content, SFD naturally gives rise to \emph{two} qualitatively different types of pre-big-bang evolutions. The first one is obtained from scale factor inversions combined with conformal time \emph{translations}, and  it consists of a collapsing universe which ends in a decelerated phase leading to a big-crunch identified with the big-bang of the subsequent expanding evolution. This is similar in spirit to the ekpyrotic cosmologies \cite{steinhardt2002cosmic,turok-seiberg} and to the original Veneziano E-frame pre-big-bang models \cite{venez-contr}. In the second possibility, scale factor inversion is  combined  with conformal time \emph{reflections}. Then both the post- \emph{and} the pre-big-bang phases are \emph{expanding} universes: one ends at an accelerated expansion followed by the big-bang of the other, as in the Conformal Cyclic Cosmologies of Ref.\cite{penrose}.

The construction of pre-big-bang phases derived from SFD is developed in Sect.\ref{Sect.SFDasSymm}. 
In Sect.\ref{Sect.scalarsfd} we elaborate on the effect of SFD in matter fields, in particular its description in terms of a self-interacting scalar field, and introduce the concept of duality and partial self-duality between the  matter fluids filling two consecutive aeons.
 The similarities and the differences between SFD cosmologies and CCC,  together with the restrictions needed to construct SFD symmetric models compatible with CCC, are presented in Sect.\ref{Sect.CCC}. It is also shown that the SFD symmetry, when applied to CCC models,  selects a family of cosmological models  whose  matter content  coincides with the $SU(1,1)/U(1)$  gauged K\"ahler sigma models with a specific self-interaction.
  In Sect.\ref{Sect.TDSFD} we derive the consistency conditions imposed  on the generalized second law of thermodynamics in the case of the Gauss-Bonnet extension of these cyclic cosmological models.
 Our concluding Sect.\ref{Sect.conclude} is mainly devoted to the discussion of  the consequences of the scale factor duality  on the properties of the adiabatic fluctuations of the SFD symmetric FRW backgrounds, with also some comments on the holographic features of the conformal crossover.



\section{Pre-big-bang evolution from SFD} \label{Sect.SFDasSymm}

Let $\ch a(\ch \eta)$ be the scale factor of a flat ($K=0$) expanding universe with a big-bang.  
The SFD transformations  gives us a dual universe. We shall describe now what are the allowed behaviors of 
the dual scale factor $\ha a(\ha \eta)$ with the dual conformal time $\ha \eta$.

\subsection{SFD-symmetric pre-big-bang extensions}

 The  SFD transformation of the energy density, Eq.(\ref{sfdro}), when combined with the Friedmann equations result in \footnote{The ``checks and hats'' notation adopted in this paper follows Penrose \cite{penrose}. The $\check{}$ mimics a future lightcone, and the $\hat{}$ a past lightcone, therefore their usage for post- and pre-big-bang evolutions, respectively.}
\br
\ch a \ch H = \pm \ha a \ha H . \label{pmaH}
\er
Hence the dual of an expanding universe, with $\ch H > 0$, may be either another expanding universe, with $\ha H > 0$, or a collapsing one, with $\ha H < 0$.
It is Eq.(\ref{pmaH}), together with the inversion of the scale factor, that fixes the relation between $\ch \eta$ and $\ha \eta$ to be linear: choosing $\ch \eta = 0$ for some $\ch a_0$ and following the positions of the signs, we have
\br
\ch \eta = \int_{\ch a_0}^{\ch a} \frac{d \ch a}{\ch a} \, \frac{1}{\ch a \ch H} = \mp \int_{c_0^2/\ch a_0}^{\ha a} \frac{d \ha a}{\ha a} \, \frac{1}{\ha a \ha H} = \mp \ha \eta + {\text{constant}} . \label{chetahaetaislinear}
\er
Thus we arrive at the conclusion that,  
for a given \emph{post-big-bang universe}, the scale factor inversions give rise to \emph{two kinds of pre-big-bang  extensions}: (i)  pairs of expanding/expanding (henceforth abbreviated as `E/E') dual universes that are related by a conformal time reflection plus a translation and (ii) pairs of contracting/expanding (`C/E') dual universes,  related by a simple time translation.

\begin{figure}[ht] 
\centering
\subfigure[]{
\includegraphics[scale=0.6]{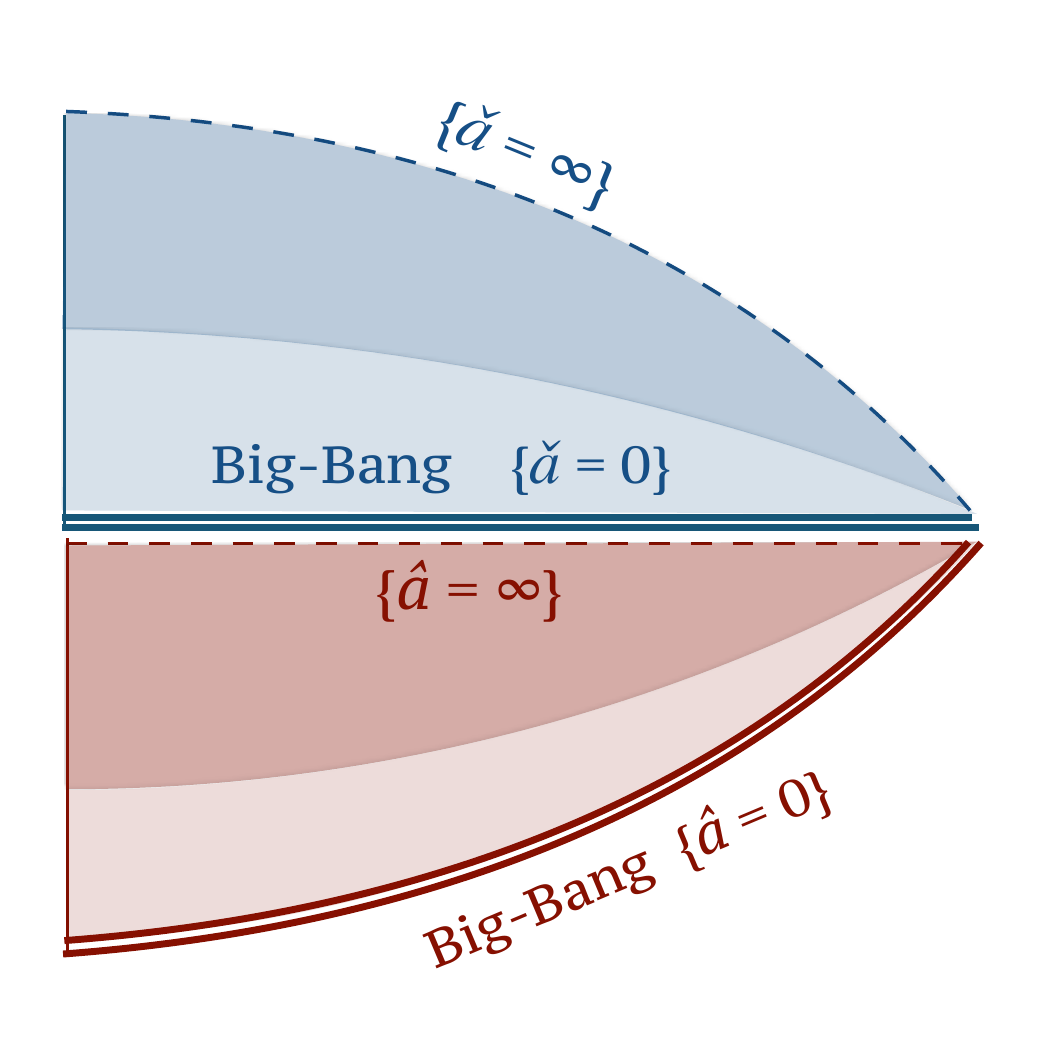}
\label{DualCCC}
}
\subfigure[]{
\includegraphics[scale=0.6]{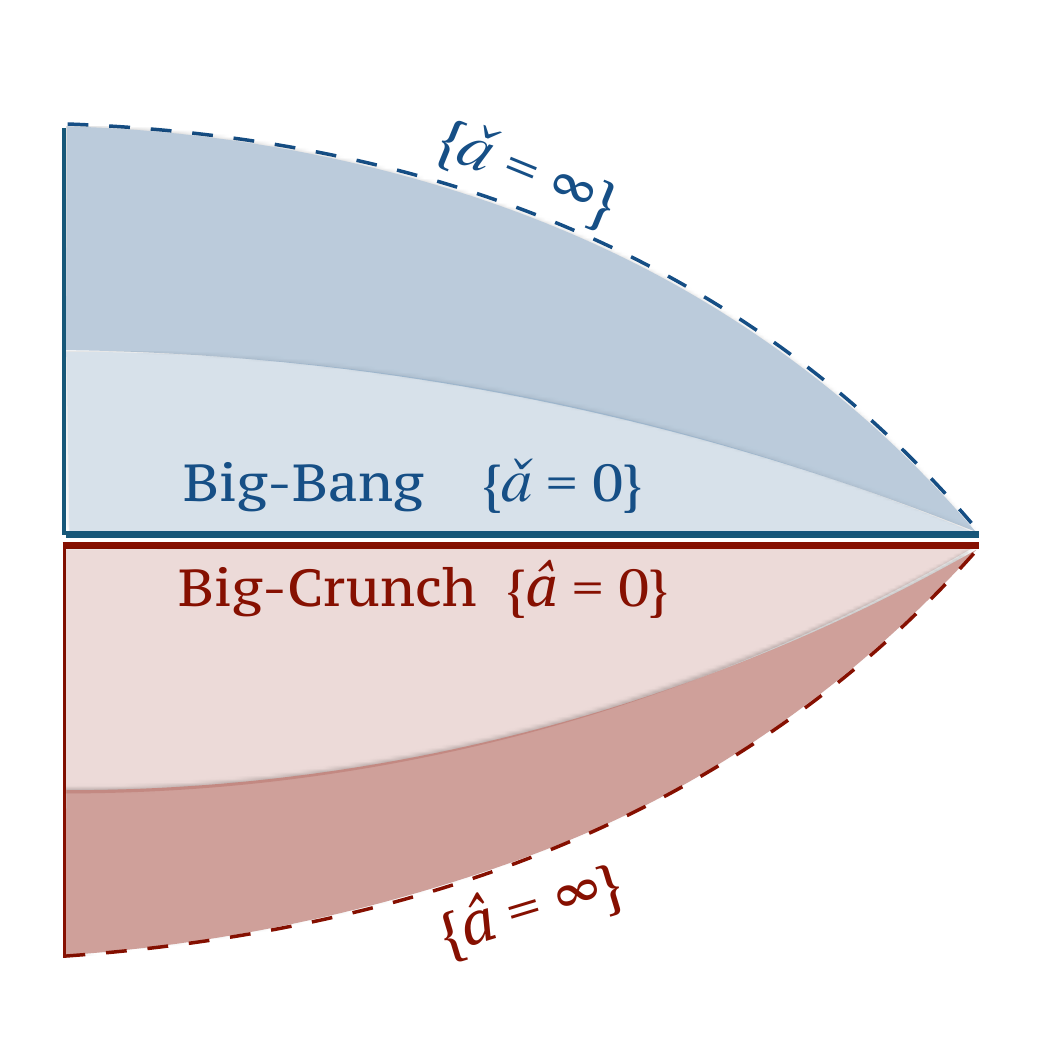} \label{DualBBBC}
}
\caption{Conformal diagram for two consecutive dual universes with two periods of acceleration; (a) expanding/expanding and (b) contracting/expanding.
Red and blue portions depict pre- and post-big-bang phases, respectively. Singularities are indicated by double lines, and infinities by dashed lines. Acceleration is positive in dark patches and negative in light ones.
}
\label{ConfDiagramsCCCBBBC}
\end{figure}

It is convenient  to place the big-bang, $\{\ch a = 0\}$, at $\ch \eta = 0$. Then
in the case of E/E  universes, Eq.(\ref{chetahaetaislinear})  becomes $\ch \eta =  - \ha \eta$ and one can further drop the marks on the $\eta$'s to write the scale factor inversion (\ref{sfinv}) as
\br
\quad \ch a \ch H = \ha a \ha H \; , \quad   \ch a ( \eta) = \frac{c_0^2 }{ \ha a(-\eta) }. \label{ExpanExpanDuali}
\er 
For a pair of C/E universes, the contracting one does not have a big-bang. Instead, it begins with $\ha a = \infty$ (at a de Sitter ``vacuum'')  and collapses to a big-crunch.
We may place both singularities at $\ch\eta = \ha\eta = 0$, for
\br
\ch \eta = \int_{0}^{\ch a} \frac{d \ch a}{\ch a^2 \ch H}  =  \int_{0}^{\ha a} \frac{d \ha a}{\ha a^2 \ha H} - \int_0^\infty  \frac{d \ha a}{\ha a^2 \ha H} = \ha \eta + \eta_f , \nonumber
\er
where $\eta_f$ is the conformal time duration of the universe:
\br
 \quad  \eta_f \equiv \int_\infty^0  \frac{d \ha a}{\ha a^2 \ha H} = \int_0^\infty  \frac{d \ch a}{\ch a^2 \ch H}. \label{etafdualetaf}
\er
So here the scale factor inversion (\ref{sfinv}) takes a form
\br
\quad \ch a \ch H = - \ha a \ha H \; , \quad   \ch a ( \eta) = \frac{c_0^2 }{ \ha a(\eta -\eta_f) }.\label{ContraExpanDuali}
\er 
A  late-time acceleration of the post-big-bang universe, as it happens in the $\Lambda$CDM model, renders the conformal time duration  $\eta_f$ \emph{finite}. As a consequence of Eq.(\ref{etafdualetaf}), the corresponding dual universe also has \emph{the same} finite conformal lifetime, and begins at $-\eta_f$ with $\ha a(-\eta_f) = 0$ or $\ha a (-\eta_f) = \infty$ in the expanding or contracting cases, respectively
\footnote{As mentioned above, SFD maps accelerated into decelerated phases. Thus for a C/E pair, the phases will be mapped in ``reversed order'', i.e. one universe has a late-time and the other an early-time accelerated phase. This is depicted in Fig.\ref{ConfDiagramsCCCBBBC}.}.
 This provides the following picture.
As the time parameter $\eta$ evolves, what we get is a whole history of an expanding or contracting universe $\{\ha a, \ha\rho, \ha p\}$ for $\eta \in [-\eta_f, 0]$ which is then \emph{followed} by a big-bang and by a dual universe $\{\ch a, \ch \rho, \ch p\}$ evolution  as $\eta$ runs forward in $[0, \eta_f]$.
Following Penrose \cite{penrose}, we will call  each of these two ``consecutive cosmological histories'' an \emph{aeon}.\footnote{Note that, although ``consecutive'',  both histories are \emph{eternal} in cosmic time, e.g. as $\eta \to 0-$, cosmic time diverges, $\ha t \to \infty$. Thus the name ``aeon''.}


\subsection{Dual Conformal Diagrams}	\label{subsectDualConfDiagrams}

The effect of the SFD transformation in the conformal diagrams sheds some insight into how the causal structures of dual universes are related.
The Penrose diagram of a flat FRW universe is obtained by the
usual coordinate transformation $(\eta, r) \mapsto (\tau , \chi)$,
\bsub\label{conftransfetarchitau}\br
& \eta  = \frac{1}{2} \sec \left(\frac{\chi + \tau}{2}\right) \, \sec \left( \frac{\chi - \tau}{2} \right)  \, \sin \, \tau \; ; \label{conftransfetarchitaua}\\
& r =  \frac{1}{2} \sec \left(\frac{\chi + \tau}{2}\right) \, \sec \left( \frac{\chi - \tau}{2} \right)  \, \sin \, \chi \; ,  \label{conftransfetarchitaub}
\er\esub
which takes the (flat) FRW line element into
\br
& d s^2 =\frac{1}{4} \, a^2(\eta) \sec^2 \left( \frac{\chi + \tau}{2} \right) \, \sec^2 \left( \frac{\chi - \tau}{2} \right) \;  ds_{Einst}^2  ,\label{einsteinconfmetric}
\er
where
$ds_{Einst}^2 = - d\tau^2 + d\chi^2 + \sin^2 \chi \left( d\theta^2 +  \sin^2\theta \, d\varphi^2 \right)$ 
is the metric of the Einstein universe (a 4-cylinder). The patch on the Einstein cylinder covered by a given FRW universe is bordered by the  surfaces corresponding  to the zeroes and infinities of the conformal factor multiplying $ds_{Einst}^2$ above.

If we choose the origin of conformal time such that the big-bang is placed at $\eta = 0$, we see that the SFD transformation (\ref{ExpanExpanDuali}) relating E/E universes, $\ha \eta = - \ch \eta$, is equivalent to $\ha \tau = - \ch \tau$. Also, the inversion of the scale factor maps singularities into conformal infinities and vice-versa. The overall effect is that, given a Penrose diagram, its dual will be the mirrored image over the invariant line $\ch \tau = \ha \tau = 0$, with the roles of the lines corresponding to singularities and infinities interchanged. 
This can be seen, e.g. in Fig.\ref{DualCCC}.
Thus we map initial into final conditions for the Friedmann equations, and vice-versa. 
The null-cone structure is, of course, mapped into itself. But note that a future null-cone in a universe is mapped into a past null-cone in its dual. In particular, this means that the particle and event horizons are mapped into each other (c.f. \cite{camara}).
This fact may be derived explicitly by writing the particle and event horizon radii, respectively,
 $\eta_p( a) = \int_0^{ a} \frac{d  a}{ a^2 H}$ and $\eta_h( a) = \int_{ a}^\infty \frac{d  a}{ a^2 H }$, as functions of the scale fator, then using Eq.(\ref{ExpanExpanDuali})  to find
\br
\ch \eta_p = \ha \eta_h \, , \quad {\text{and}} \quad \ch \eta_h = \ha \eta_p. \label{phetaCCClike}
\er
An interesting consequence of Eqs.(\ref{phetaCCClike}) is to show how if the big-bang at $\{ \ch a = 0\}$ is space-like --- a characteristic of decelerated expansion ---, then so is the future infinity $\{\ha a = \infty\}$ of the dual universe --- a characteristic of accelerated expansion, as it was to be expected. 


\begin{figure}[ht] 
\centering
\subfigure[]{
\includegraphics[scale=0.6]{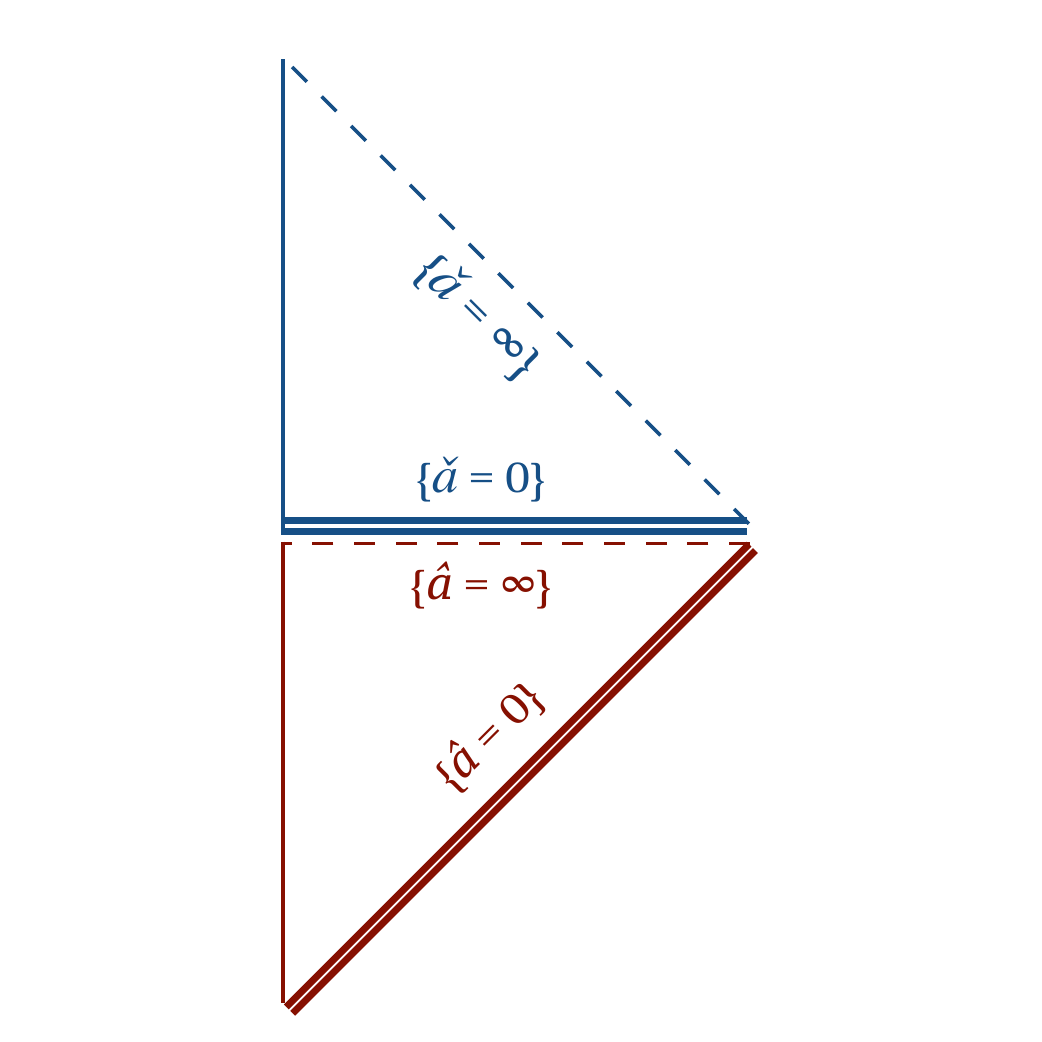}
\label{PerfectFluidsPenrCCC}
}
\subfigure[]{
\includegraphics[scale=0.6]{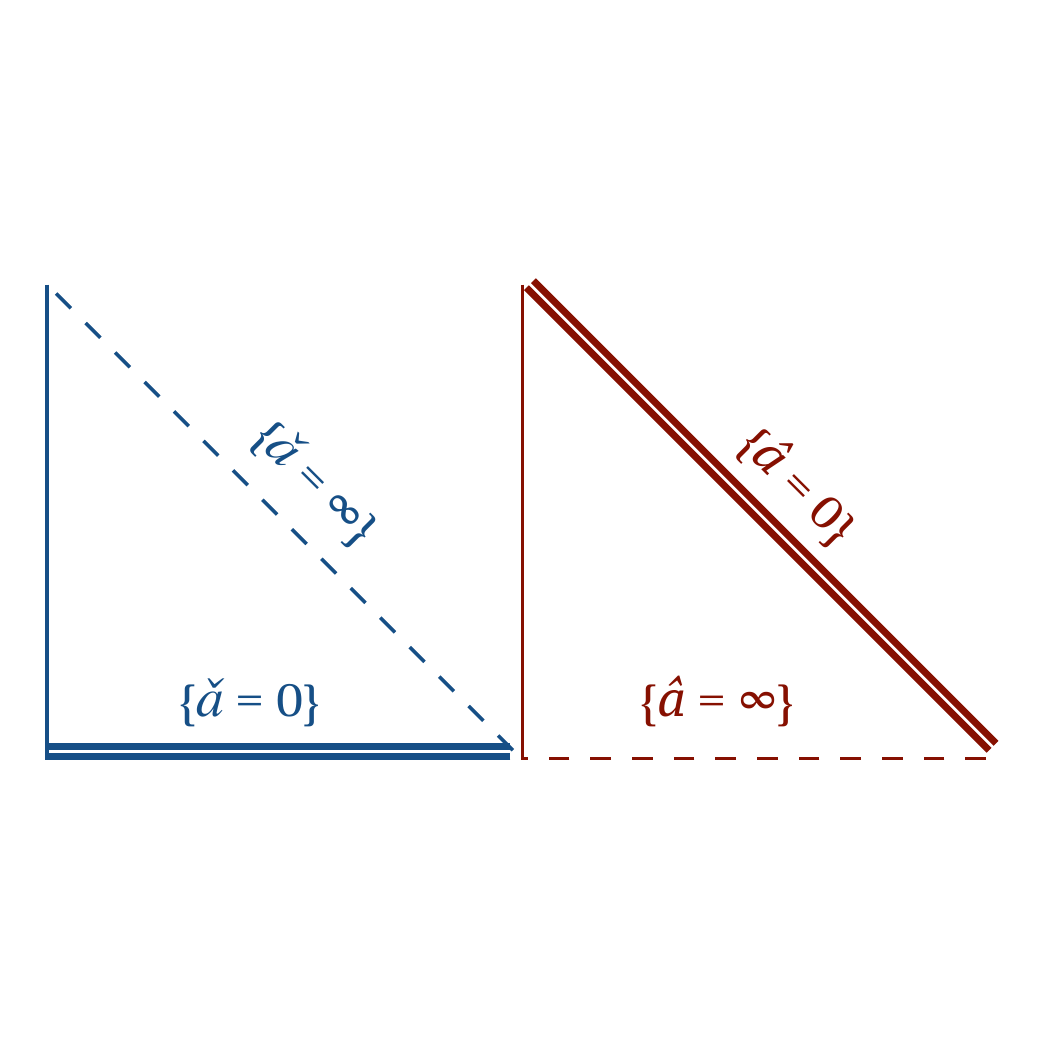}
\label{PerfectFluidsPenrBBBC}
}
\caption{Causal diagrams for dual universes filled with fluids with constant equation of  state parameters.  The blue diagrams represent decelerated expansion ($\ch w > - \frac{1}{3}$) and have only a particle horizon. The dual red diagrams are both accelerating ($\ha w < - \frac{1}{3}$).
(a) Expansion/Expansion: the dual red diagram has accelerating expansion and  only an event horizon. (b) Expansion/Contraction: the red diagram has accelerated contraction and  only a particle horizon.}
\label{PerfectFluidsPenr}
\end{figure}


For a pair of C/E dual universes we do not have the $\tau \mapsto - \tau$ mapping, and thus there is no ``flipping'' of the conformal diagram. We may, however, as has been noted above, choose the origin of conformal times such that conformal infinity and the dual big-bang coincide. The scale factor inversion, nevertheless, still exchanges the roles of these asymptotic surfaces. The overall effect for a universe with a finite conformal lifetime can be seen in Fig.\ref{DualBBBC}. As to what regards the horizons, we have now:  $\eta_p = \int_\infty^a da (a^2 H)^{-1}$ and $\eta_h = \int_a^0 da (a^2 H)^{-1}$, which leads us to conclude that $ \ch \eta_p =  \ha \eta_p$.
A similar calculation is valid for event horizons $\ch \eta_h = \ha \eta_h$. 
Therefore the particle horizons of the expanding and the contracting universes are mapped into each other, and so are the respective event horizons.

We shall be  concerned in this paper  with  universes like the ones appearing in Fig.\ref{ConfDiagramsCCCBBBC},  in which there are two different phases of  acceleration (such as in the $\Lambda$CDM model), thus possessing \emph{both} kinds of horizon and marked by a \emph{finite  total conformal time duration} $\eta_f = \int_0^\infty  d  a \ ( a^2  H)^{-1} =  \eta_p +  \eta_h$.  
This may, of course,  not be always the case. For example,  universes filled with a single perfect fluid with constant equation of state parameter $\ch w > - \frac{1}{3}$, and their duals, with  $\ha w < - \frac{1}{3}$, have, each, only one kind of horizon as shown in Fig.\ref{PerfectFluidsPenr}. Then either $\eta_p$ or $\eta_h$ are infinite, resulting in an infinite $\eta_f$ so Eq.(\ref{ContraExpanDuali}), for example, has to be modified. 
The crucial point here is this:
When both boundary surfaces $\{a = 0\}$ and $\{a = \infty\}$ are space-like, as in Fig.\ref{ConfDiagramsCCCBBBC}, we may \emph{identify} the coinciding boundaries of consecutive aeons as \emph{one single space-like hypersurface} $\cal X$, which we shall call a \emph{`crossover surface'}. It is conformally equivalent to the Euclidean 3-dimensional space (that is, a constant-time-surface in Minkowski space-time), and with our choice of conformal time, $\cal X = \{ \eta = 0\}$. Now in the case of Fig.\ref{PerfectFluidsPenr}, one cannot define properly a transition surface because this would amount to identify hypersurfaces with different signatures.  For example, in a C/E pair the contracting universe ends in a null hypersurface while the expanding universe begins in a space-like big-bang (Fig.\ref{PerfectFluidsPenrBBBC}). As for the transition in a E/E pair, it may be well defined but it can be done only once; that is, it existis only for ``two consecutive aeons'', and one cannot naturally build a ``chain'' of aeons, as it will be done presently in Sect.\ref{CyclciSFDextensions}.

\subsection{Cyclic SFD extensions} \label{CyclciSFDextensions}

\begin{figure}[ht] 
\centering
\subfigure[]{
\includegraphics[scale=0.5]{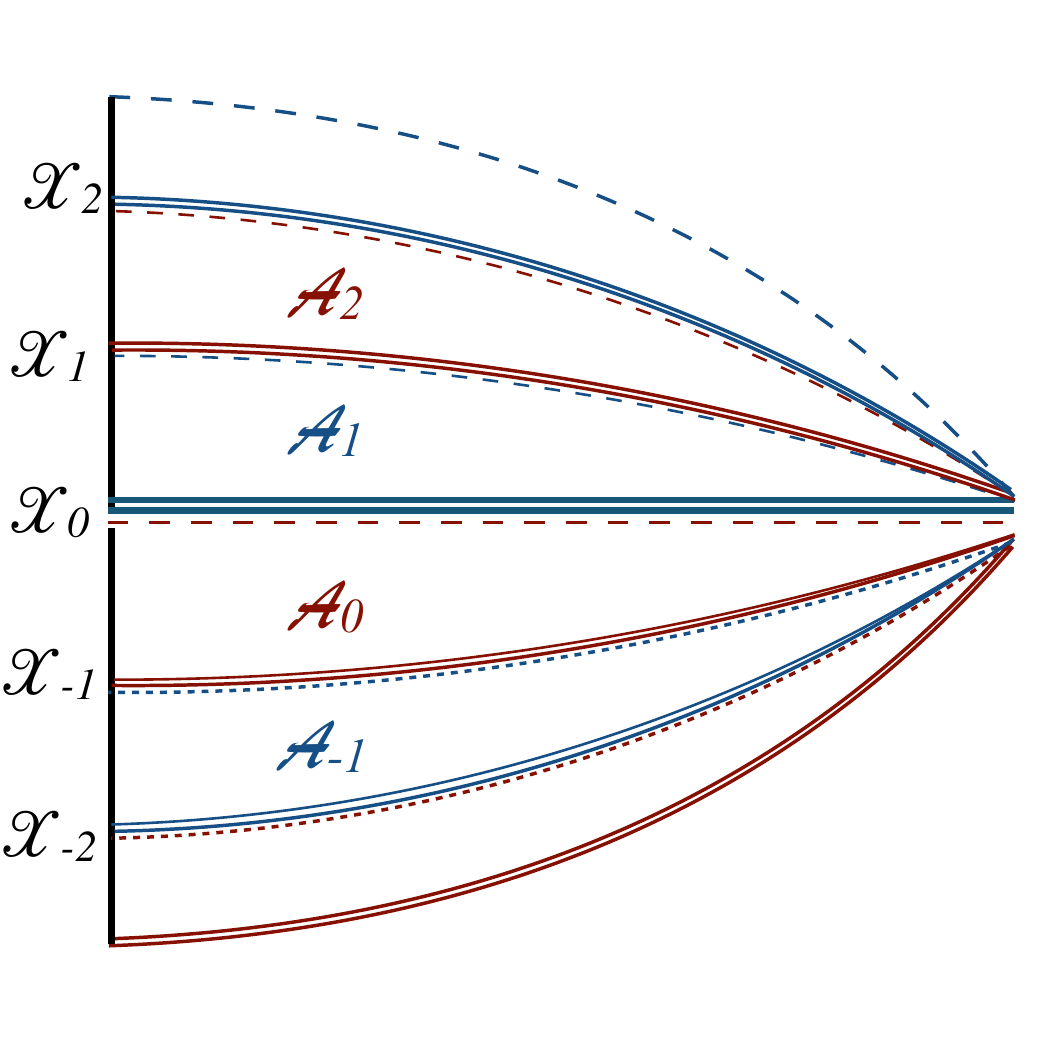} \label{AeonsChainPenrose}
}
\subfigure[]{
\includegraphics[scale=0.5]{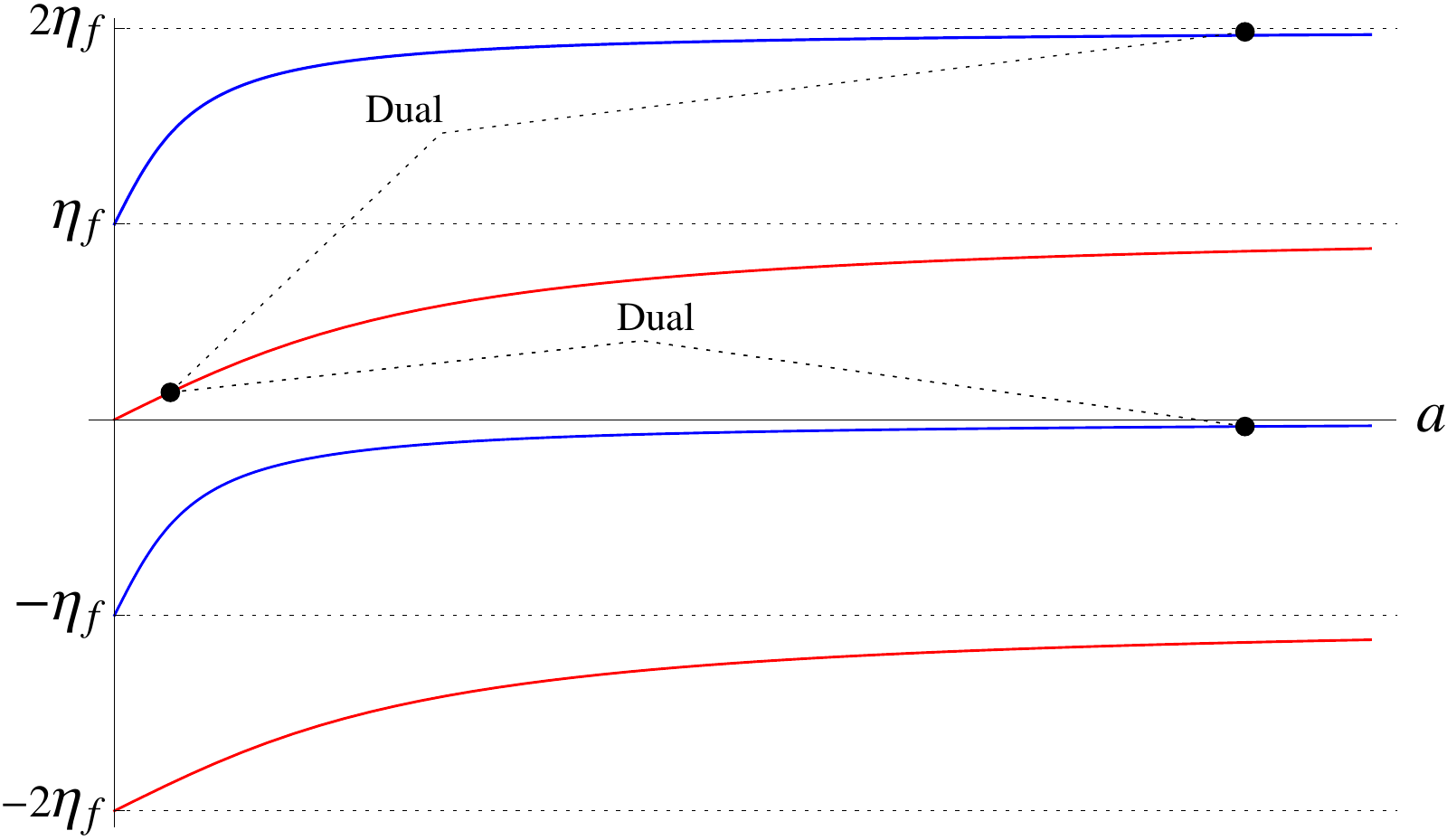} \label{RecicledCosmology}
}
\caption{Sequences of aeons. Same colors indicate aeons with identical matter and scale factor. (a) Conformal diagram representing the sequence with its space-like crossover surfaces $\cal X_j$; double lines indicate a singularity ($a_j = 0$), and dashed lines a conformal infinity ($a_j = \infty$). (b) The conformal time evolution of the scale factors, as given implicitly by $\eta_j(a_j)$. The black dots mark a sequence of dual instants  related by SFD according to Eq.(\ref{SFIrecurs}).
}\label{RecicledCosmologyAll}
\end{figure}

The rather unusual realization of the transition between expanding/expanding SFD universes 
--- with the future infinity of one aeon being identified with the initial big-bang of the next ---
suggests that  the pattern should repeat itself indefinitely. That is, it is natural to define a  \emph{sequence} of aeons, each being related to its following neighbor by an SFD transformation.
We denote by 
\br
\cal A_j = \{ a_j (\eta_j) ; \quad \eta_j \in [ (j-1) \eta_f , j \eta_f ] ; \quad j \in {\mathbb Z} \}		\label{AeonjDef}
\er
each aeon in the sequence. See Fig.\ref{RecicledCosmologyAll}.
We thus partition the real line $\eta \in \mathbb R$ into a set of intervals $[ (j-1) \eta_f , j \eta_f ]$, with the same length, which are the domains of the scale factors $a_j(\eta_j)$: we have  $a_j (j \eta_f) = \infty$ and $a_{j} \left( (j-1) \eta_f \right) = 0$. Note that the order of the indices  $j$ follow the positions of the  ``ends of the universe",  e.g. $\cal A_2$ terminates at $\eta = 2\eta_f$, $\cal A_1$ at $\eta = 1 \eta_f$, $\cal A_0$ at $\eta = 0$, etc. 
Note also that, by construction and as it should be, according to Eq.(\ref{etafdualetaf}), every aeon $\cal A_j$ in the sequence has the same  conformal ``lifespan" $\eta_f$.

Our construction depends upon $\eta_f$ being finite. As discussed in the preceding Sect.\ref{subsectDualConfDiagrams}, this implies that the surfaces $\cal X_j \equiv \{\eta = j \eta_f\}$ are 3-dimensional space-like crossover surfaces between the aeons. Alternativelly, 
$$\cal {X}_j  \equiv \big\{ a_j (j \eta_f) = \infty \approx a_{j+1} \left( j \eta_f \right) = 0\big\}.$$
See Fig.\ref{AeonsChainPenrose}.

%
%
The scale factor $a_j(\eta_j)$ of each aeon determines the scale factor of the next one via the SFD transformation, which reads
\br
\quad a_{j+1} (\eta_{j+1}) = \frac{c_0^2}{a_j ( - \eta_{j+1} + 2j \eta_f)} \; ; \;\; a_{j-1}(\eta_{j-1}) = \frac{c_0^2}{a_j (-\eta_{j-1} +2(j-1)\eta_f)} .
 \label{SFIrecurs}
\er
Accordingly, the densities and pressures of the fluids are related  by  the periodic analogs of Eqs. (\ref{sfinv}), (\ref{sfdro}) and (\ref{sfdw}):
 \br
& a^2_j \, \rho_j = a^2_{j+1} \, \rho_{j+1} , \quad  a^2_j ( 3 p_j+ \rho_j ) = - a^2_{j+1} ( 3p_{j+1}+\rho_{j+1} ) ,\nonumber\\
 & w_j + w_{j+1} = -\tfrac{2}{3}.
 \label{sfdciclic}
\er
Eqs.(\ref{AeonjDef})-(\ref{sfdciclic}) define the sequence of aeons we have proposed.  Regardless of the infinite number of aeons, this cyclic SFD model  consists of a repetition of  the \emph{same} dual pair $\{a_j, a_{j+1}=c_0^2/a_j\}$. Indeed, it is
 an important consequence of  the ``cyclic $Z_2$ form''  of the scale factor inversions in (\ref{SFIrecurs}), that  $ a_{j-1}=a_{j+1}$, as can be seen in Fig.\ref{RecicledCosmology}. Naturally, the matter content presents analogous relations, with $\rho_{j-1}=\rho_{j+1}$ and $p_{j-1}=p_{j+1}$, while $\rho_j$ is related to as  $\rho_{j+1}=\tfrac{a^4_j}{c_0^4} \rho_j$, etc.  An explicit example of our construction  is given in App.\ref{AppA}.

\section{Scale factor duality for matter fields} \label{Sect.scalarsfd}

The matter content of the SFD-symmetric  pre-big-bang cosmologies consists of pairs of barotropic fluids or, alternatively, pairs of self-interacting scalar fields. The pre- and post-big-bang equations of state of the fluids, as well as the corresponding scalar field potentials, are related by the transformations (\ref{sfdro}).  In this section we investigate  the most general consequences and restrictions imposed  by these extension rules  on  the  matter  constituents. We are mainly interested on  their interrelations, such as  $\ha\rho(\ch\rho)$ and   $\ha\s (\ch\s)$.

\subsection{Dual and Partially  Self-Dual Fluids} \label{SubSectDualPatDual}

 According to the requirement of  SFD symmetry,  for each  given post-big-bang density $\ch\rho(\ch a)$, one can   easily  determine the form of its  dual, 
 \br
\ha\rho (\ha a ; \ha\rho_J)=\tfrac{1}{c_0^4} \, \ch a^4 \; \ch\rho (\ch a; \ch\rho_J) . \label{ha}
\er
We have introduced explicitly here the collections of relative densities parameters $\ch \rho_J$ and $\ha \rho_J$,  because (\ref{ha}) implies relations between them and the SFD parameter $c_0$. For example, the $\La$CDM parameters, $\rho_J \equiv \{ \rho_r, \rho_d, \rho_\La\}$,  relate to their dual as in Eq.(\ref{densreladualLACDM}).  
We shall  consider a special class of cosmological models satisfying an additional \emph{partial self-duality condition}\footnote{It is similar (but not identical) to the partial self-duality concept introduced  in Sect.3 of Ref.\cite{selfdual}, within the framework of  3D New Massive Gravity Holography.} by requiring that the forms of the fluid densities and pressures (and of their EoS) are preserved by  SFD transformations, but allowing  changes  in the \emph{values} of  the fluid's parameters  $\ha\rho_J \neq \ch\rho_J$.
For example, a universe with radiation and cosmological constant  is partially self-dual since, if $\ch \rho = \ch \rho_r \ch a^{-4} + \ch \rho_\La$, then from (\ref{ha}) the dual aeon has the same constituents: $\ha \rho = \ha \rho_\La + \ha \rho_r \ha a^{-4}$, but with different relative densities $\ch\rho_J$ and $\ha\rho_J$, which are  related by $\ch \rho_r / \ha \rho_\La = c_0^4 = \ha \rho_r / \ch \rho_\La$.
We note that partial self-duality is a weaker version of the more restrictive condition of (complete) \emph{self-duality} introduced in \cite{camara}. The latter is realized by further demanding $\ha\rho_J = \ch\rho_J$, in which case both aeons are completely identical.\footnote{E.g. the only self-dual perfect fluid, with a constant EoS parameter $w$, is  the string gas model with $w= -1/3$ and $\rho(a)=\rho_{str}/a^2$. For  a self-dual composition of two fluids,  see  Ref.\cite{camara}.}

A particularly relevant family of two interacting fluids for pre-big-bang extensions is given by the modified Chaplygin gas model \cite{chap-modif1, Benaoum:hh,chap-mod2}
\br
\rho=\Big( \rho_\La^\delta +  \rho_r^\delta \; a^{-4\delta}\Big)^{\frac{1}{\delta}},\quad\quad p=\tfrac{1}{3} \rho -\tfrac{4\rho_\La^\delta}{3}  \;  \rho^{1-\delta},\quad\quad  0 < \delta \leq 1.
\label{chaply}
\er
We  have chosen to study the cyclic extensions for a subclass of models  determined by the restriction $0 < \delta \leq 1$.  The upper limit for $\delta$ comes from the  requirement  that the corresponding curvatures $R(\delta)=2 \rho^{\delta}_{\Lambda} \left( \rho \right)^{1-\delta}$ are monotonically decreasing functions of the density $ \infty \geq\rho\geq\rho_{\Lambda}$. The lower limit $\delta>0$ selects matter fluids which give  rise to a finite conformal time lifespan, $\eta\in(0,\eta_f)$, and  behave  asymptotically as a cosmological constant for $a \rightarrow\infty$ and  as radiation when $a\rightarrow 0$. By construction they imply the existence of big-bang singularity followed by a  decelerated   and then an accelerated  period of expansion in the post-big-bang phase.\footnote{Notice that one of its  SFD dual geometries describe a contracting  universe with monotonically increasing curvature starting from the initial de Sitter state of  $\rho_{\Lambda}$ and reaching the bing-crunch singularity at $\rho\rightarrow\infty$.} Acceleration changes sign (i.e. vanishes) at a critical value of the scale factor, $ a  =  a_{cr}$, or, alternatively, a critical value of the energy density, $ \rho =  \rho_{cr}$, such that
\br
 a_{cr} = \left(\frac{ \rho_r}{ \rho_\La}\right)^{\frac{1}{4}} ,	\quad \quad	 \rho_{cr} = 2  \rho_\La ,	\label{critpoint}
\er
when the deceleration parameter $q(a)\equiv\frac{1}{2}\left(1+3\frac{p}{\rho}\right)$ vanishes. 
The modified Chaplygin models are in fact the only ones to behave asymptotically as a cosmological constant within a very general class of partially self-dual fluids \cite{camara}, which makes them particularly useful in the construction of SFD symmetric CCC models, realized in  Sect.\ref{Sect.CCC} below. Their conformal lifetime is given by:
\br
\eta_f=\frac{\sqrt{3}}{4\delta \, (\rho_r \rho_\La)^{\frac{1}{4}}} \frac{\left[\Gamma\left(\frac{1}{4\delta}\right)\right]^2}{\Gamma\left(\frac{1}{2\delta}\right)}, \label{aetafal}
\er
as one can easily confirm from the analytic form \cite{camara} of the (implicit) scale factor $\eta = \eta(a)$ evolution. Details concerning a particular example of $\delta=1/2$ model  and its cyclic extension are  presented in App.\ref{AppA}.

The relation between the parameters $\rho_J = \{ \rho_r, \rho_\La\}$ for two dual aeons is fixed by (\ref{ha}) to be
\br
c_0^{4}=\ch \rho_r / \ha \rho_\La = \ha \rho_r/ \ch \rho_\La, \label{rhochanges}
\er 
which keeps invariant (\ref{aetafal}); i.e. $\ch \eta_f = \ha \eta_f = \eta_f$ as was to be expected.
The dual pre-big-bang evolution also contains both periods, with the SFD transformations mapping accelerated into decelerated phases and vice-versa, as discussed in Sect.\ref{Sect.Intro}.
The simple form the fluid density (\ref{chaply}) allows us to  derive  from Eq.(\ref{ha})  a  relation between the  post- and pre-big-bang energy densities:
\begin{eqnarray}
\ha {\rho}^{\delta}-\ha\rho_{\Lambda}^{\delta}=\frac{\ha\rho_{\Lambda}^{\delta} \ch\rho_{\Lambda}^{\delta}}{\ch\rho^{\delta}-\ch\rho_{\Lambda}^{\delta}}, \label{daulrho}
\end{eqnarray} 
   which shows explicitly how the  SFD transformations act on the matter content exchanging initial and final conditions. 
 For example, the duality between the scale factors  $\ch a=0$ and $\ha a=\infty$ at $\ha\eta=0=\ch\eta$   relates  $\ch\rho(0)=\infty$  to  $\ha\rho(\infty)=\ha\rho_{\Lambda}$ and vice-versa.
 The above high-to-low  densities  transformation%
 \footnote{Together  with a similar duality relation for the curvatures, viz. $\ha R=2 \ha\rho^{\delta}_{\Lambda} \left( \ha\rho \right)^{1-\delta}$ and $\ch R=2 \ch\rho^{\delta}_{\Lambda} \left( \ch\rho \right)^{1-\delta}$.}  
 highlights the intrinsic UV/IR nature of  conformal time scale factor duality.

 We have to also mention another important feature of every partially self-dual SFD model, and in particular of  the modified Chaplygin gas (\ref{chaply}), namely their invariance under another conformal SFD transformation, acting now \emph{within} each aeon $\cal A_j$, that maps  the early-time decelerated phase into the late-time accelerated phase as follows:
\be
\til a_j(\eta_j)=\frac{c_{0c}^2(j)}{a_j(\eta_f - \eta_j)}, \quad\quad c^4_{0c}(j)=\frac{\rho^{(j)}_r}{\rho^{(j)}_{\Lambda}}=\left(\frac{\rho^{(j)}_d}{\rho^{(j)}_{dw}} \right)^2 . \label{sfdca}
\ee
Observe that here we have a (very specific) combination of  time reflection \emph{and} translation (with a fixed point $\eta_c=\eta_f/2$). 
The  composition of the two different SFD transformations --- one acting within each aeon with the intrinsic parameter $c^2_{0c}(j)$,  followed by the inter-aeonic SFD given by Eqs.(\ref{SFIrecurs}) and (\ref{sfdciclic})  --- defines  a   \emph{double SFD $Z_2\times Z_2$ symmetry}. It is expressed between two consecutive aeons as
\br
\til{\ch a}(\eta)=\frac{\ch c_{0c}^2}{\ch a(\eta_f - \eta)}=\frac{\ch c_{0c}^2}{c_0^2} \ha a(\eta-\eta_f)=\frac{\ch c_{0c}^2 \ha c_{0c}^2}{c_0^2 \til{\ha a}(-\eta)}=\frac{c_0^2}{\til{\ha a} (-\eta)} .
 \label{doublesfd}
\er
The identity $c^4_0=\ch c^2_{0c} \ha c^2_{0c}$, automatically satisfied by the parameters $c^4_0=\ha\rho_r/\ch\rho_{\Lambda}=\ch\rho_r/\ha\rho_{\Lambda}$, $\ch c^2_{0c}=\sqrt{\frac{\ch\rho_r}{\ch\rho_{\Lambda}}}$ and $\ha c^2_{0c}=\sqrt{\frac{\ha\rho_{\Lambda}}{\ha\rho_r}}$, ensures the consistency of these double SFD transformations.

\vspace{0.3cm}

\subsection{Dual Scalar Fields}
 
We next consider the equivalent  scalar field description of the pairs of dual  fluids,  and the SFD transformations between them. Each of the pre- and post-big-bang  fluids can be replaced by corresponding scalar fields $\ha \s$ and $\ch \s$, with potentials $\ch V(\ch \s; \ch\rho_J)$ and $\ha V(\ha \s;  \ha\rho_J)$, such that
\br
\ch \rho=\frac{1}{2} (\ch \s'/\ch a)^2+\ch V(\ch \s), \quad\quad  \ch p=\frac{1}{2}(\ch\s'/\ch a)^2-\ch V(\ch \s) ,		\label{reconstr}
\er
and the same for $\ha \s$, $\ha V(\ha \s)$, etc.
 Then the SFD transformations (\ref{sfdro}) can be rewritten as 
\br 
 \ha V(\ha\s)= \frac{1}{3} \ch V(\ch \s) + \frac{2}{3} (\ch \s'/\ch a)^2, \quad\quad (\ha \s'/\ha a)^2=\frac{4}{3} \ch V(\ch \s) - \frac{1}{3} (\ch \s'/\ch a)^2 .		 \label{scaltr}
\er
Given the potential $\ch V(\ch\s ; \ch\rho_J)$ of the post-big-bang model, our goal is to find the explicit form of the scalar fields  duality transformation  $\ha\s=\ha \s(\ch\s)$ and to use it further in the derivation of the corresponding dual or partially self-dual pre-big-bang potential $\ha V(\ha\s ; \ha\rho_J)$. 
The Friedmann equations (\ref{frw}), when combined with eqs.(\ref{reconstr}), lead to the auxiliary  relations
\br
(\ch\s')^2=-\frac{\ch a^3}{3} \frac{d\ch\rho}{d\ch a}, \quad\quad  \ch V(\ch a)=\ch\rho+\frac{\ch a}{6} \frac{d\ch\rho}{d\ch a},\quad\quad  \frac{d\ch\s}{d\ch a}=\pm \sqrt {\frac{\ch a \frac{d\ch\rho}{d\ch a}}{3K-\ch a^2\ch\rho}},\label{ident}
\er
which allow to realize  the scalar field as a function of the scale factor --- i.e.  given $\ch\rho(\ch a)$  to find  $\ch\s(\ch a)$.  In the cases  when its inverse $\ch a(\ch\s)$ can be obtained (and so  $\ch\rho(\ch\s)$ as  well),  one  derives from  eqs.(\ref{ha}), (\ref{reconstr}) and (\ref{scaltr}) the  pre-big-bang potential and the scalar fields SFD transformations:
\br
\ha V(\ha\s)=\frac{\ch a^4 (\ch\s)}{c_0^4} \Big(\frac{4}{3} \ch\rho(\ch\s) - \ch V(\ch\s) \Big) \; ; \quad \quad   \ha\s(\ch\s) = \pm \int{\sqrt{\frac{3\ch V- \ch\rho}{3(\ch\rho- \ch V)}}} d\ch\s +\s_0,  \label{dualpot}
\er
where we have also used the  scale factor inversions in the form: $\ha a (\ha\s)=c_0^2/\ch a(\ch\s)$.  A partially self-dual fluid, naturally, gives rise to a partially self-dual potential, such that $\ha V(\ha\s ; \ha\rho_J )=\ch V(\ha \s; \ha\rho_J)$.

For the modified Chaplygin gas  (\ref{chaply}), the explicit solution for the the scale factor $a(\s)$ is given by \cite{camara}
\be
\ch a(\ch\s)= \left(\frac{\ch\rho_R}{\ch\rho_{\Lambda}}\right)^{\frac{1}{4}} \left[ \sinh\left(\delta(\s_0-\ch\s)\right)\right]^{-\frac{1}{2\delta}},\label{aetaf}
\ee
with $\ch\s\leq\s_0$, and a similar one for $\ha a(\ha\s)$. Then, from eqs.(\ref{scaltr}), one can easily   find  $\ha\s(\ch\s)$,
\br 
\sinh\left(\delta (\ha\sigma- \sigma_0)\right)=\frac{1}{\sinh\left(\delta(\ch\sigma-\sigma_0)\right)}, \label{stils}
\er
and the corresponding partially self-dual potential
\begin{eqnarray} 
\ha V(\ha\sigma)=\frac{\ha\rho_{\Lambda}}{3}\Bigg\{\left[\cosh^2\left(\delta(\ha\sigma-\sigma_0)\right)\right]^{\frac{1}{\delta}}+
2 \left[\cosh^2\left(\delta (\ha\sigma-\sigma_0)\right)\right]^{\frac{1-\delta}{\delta}}\Bigg\}. \label{poten}
\end{eqnarray}
Due to the difference between the vacuum densities of the two aeons, $\ha\rho_{\Lambda}\neq\ch\rho_{\Lambda}$,  the dual fields $\ha\s$ and $\ch\s$ have different masses related by
\be
\ha m_{\ha\s}^2=\frac{\ha\rho_{\Lambda}}{\ch\rho_{\Lambda}} \ch m_{\ch\s}^2, \quad \quad \ha m^2_{\ha\s}=\ha V''(\ha\s=\s_0) = 2\delta (3 - 2\delta) \frac{\ha\rho_{\Lambda}}{3} . \label{partmassdual}
\ee


\begin{figure}[ht] 

\centering
\subfigure[]{
\includegraphics[scale=0.65]{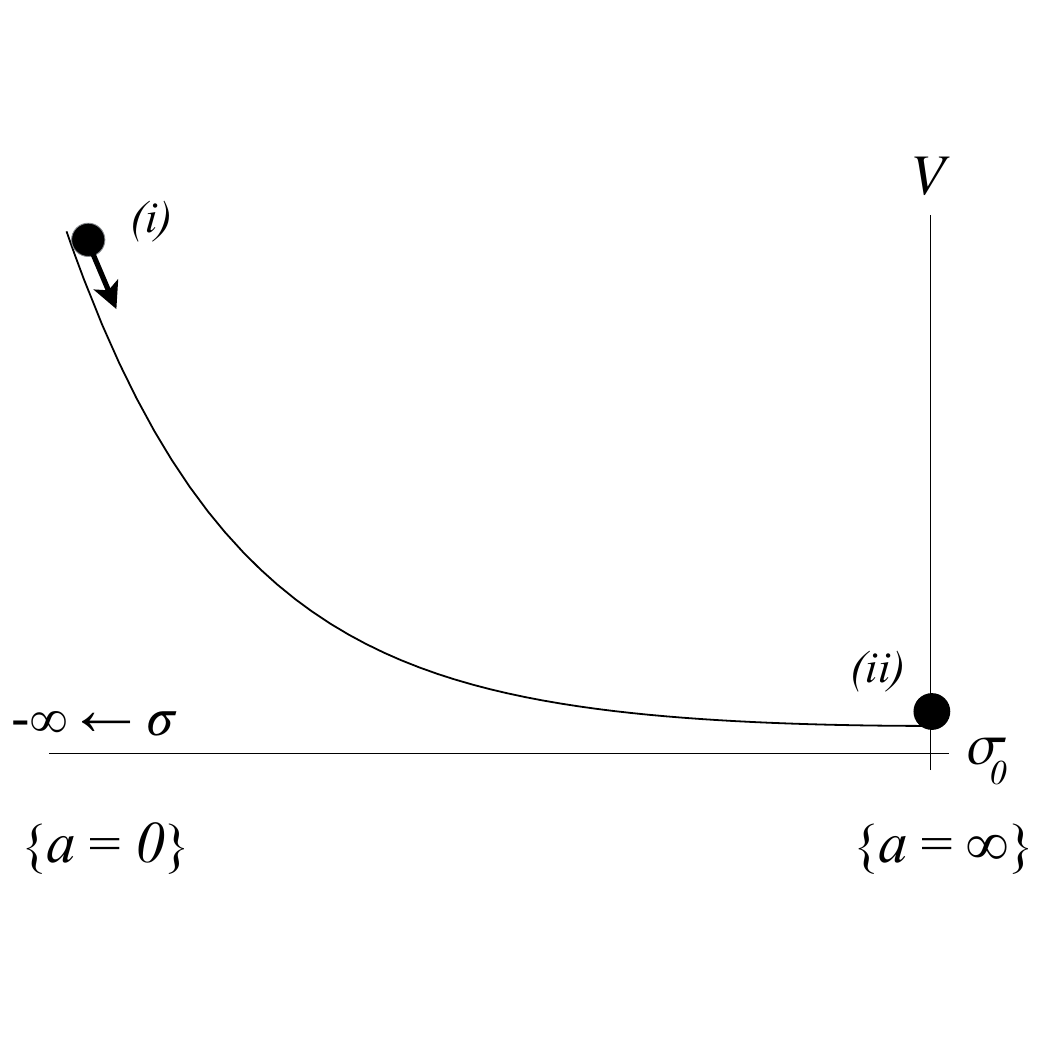} \label{VdeSigmaEE}
}
\subfigure[]{
\includegraphics[scale=0.65]{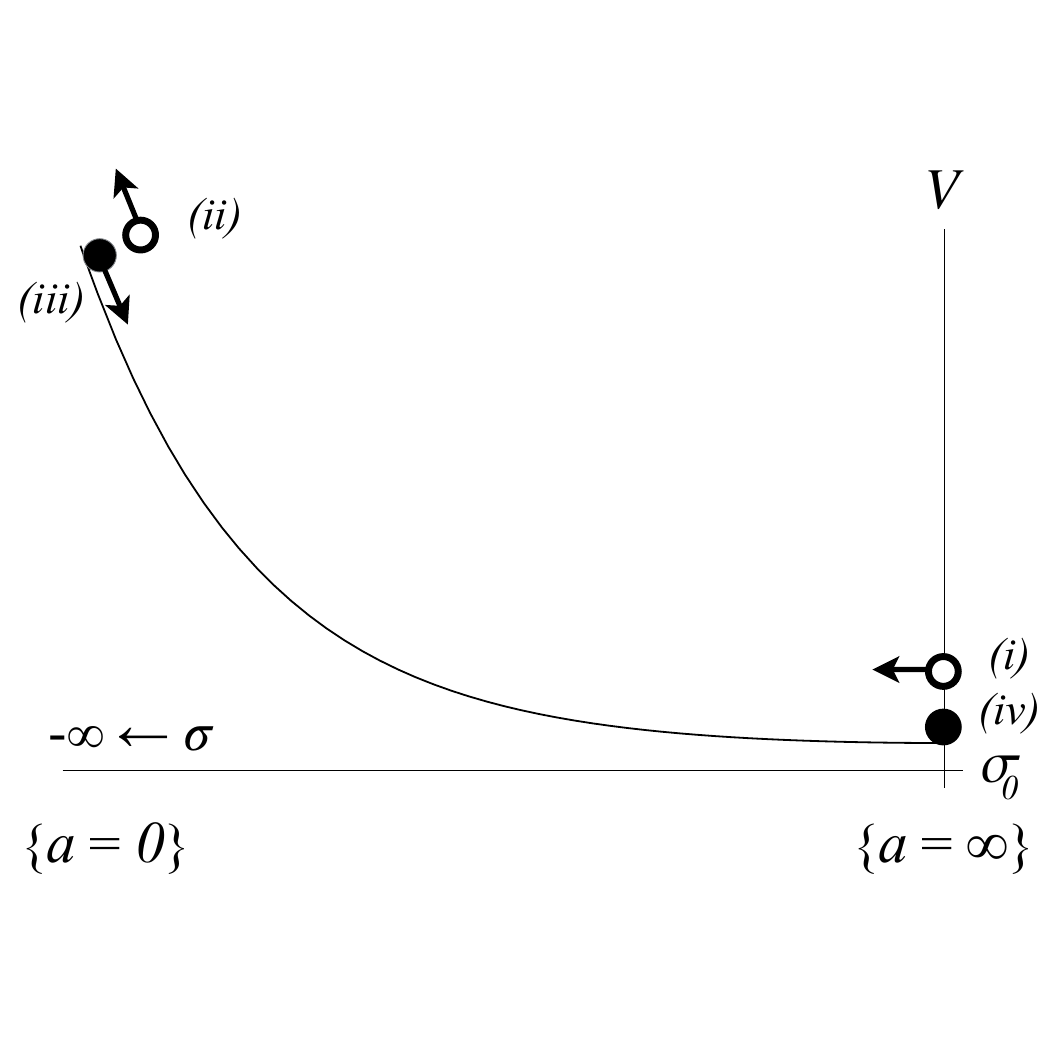} \label{VdeSigmaCE}
}
\caption{
The possible pre-big-bang evolutions of $\s$ associated with the potential (\ref{poten}). (a) E/E case: both dual fields roll down the potential towards the de Sitter vacuum. (b) C/E case: the pre-big-bang  field $\ha \s$ (white ball) begins at the vacuum and rolls up to the big-crunch; then the dual post-big-bang field $\ch \s$ (black ball) rolls down from the big-bang to de Sitter.
}\label{VdeSigma}
\end{figure}


Since the potential (\ref{poten}) is symmetric under reflections about  $\s_0$, we may choose $\s_0 = 0$ and to further consider its ``left side'' only, i.e. $\s < 0$. Then  as a consequence  Eq.(\ref{stils}), both  $\ch \s$ and $\ha \s$ will run on this chosen side.  
There is an expanding solution for the scale factor (\ref{aetaf}), with the field $\ha\s$ rolling from the big-bang at $V(\infty) = \infty$, down to the de Sitter vacuum at $V(0) = \rho_\La$. Its SFD dual  solution, obtained from Eq.(\ref{ContraExpanDuali}), describes a contracting universe, starting at the vacuum, where the field climbs the potential towards a big-crunch. Let us assume that  $\ch \s$ represents an expanding solution.  If we differentiate Eq.(\ref{stils}), we find that
\br
\frac{d \ch \s}{d \ch \eta} = - \left( \frac{\cosh \delta \ha \s}{\cosh \delta \ch \s \; \sinh^2 \delta \ha \s} \right)  \frac{d \ha \eta}{d \ch \eta} \; \frac{d \ha \s}{d \ha \eta} . \nonumber
\er
The expression in parenthesis is always positive, therefore we can have two  different cyclic extensions, of  the E/E or C/E  type, as presented in Fig.\ref{VdeSigma}. For simplicity, we now take $V(\s)$ to be ``completely self-dual'', i.e. $\ch \rho_\La = \ha \rho_\La$, so that both dual universes may be described with one single potential.

If $d \ha \eta / d \ch \eta < 0$, the field $\ha\s$ and its dual $\ch\s$  roll down\footnote{Both could also climb the potential, leading to dual contracting universes, the time-reversal of the E/E pairs. We are, as always, not considering such cases.} the potential in \emph{both} universes. This corresponds to an E/E dual pair for which, indeed, $d \ha \eta / d \ch \eta = -1$, as seen in Sect.\ref{Sect.SFDasSymm} (cf. Eq.(\ref{ExpanExpanDuali})). 
The behavior of the fields is symbolically depicted on Fig.\ref{VdeSigmaEE}. 
Here, the de Sitter vacuum  is conformally identified with the big-bang at the crossover surface, viz. $\{\ch \s = 0 \} \approx \{ \ha \s = - \infty\} \approx \cal X$. In other words, when the field $\ha \s$ reaches the de Sitter vacuum, Eq.(\ref{stils}) implies that its dual $\ch \s$ ``reappears'' at the top of the potential and yields a new big-bang and a new expanding universe as it rolls down.

The other possibility has  $d \ha \eta / d \ch \eta > 0$. This is a C/E dual pair, for which $d \ha \eta / d \ch \eta = +1$ (cf. Eq.(\ref{ContraExpanDuali})).  As depicted in Fig.\ref{VdeSigmaCE}, if $\ch \s$ rolls down, so that $d \ch \s / d \eta < 0$ and the universe expands, then $\ha \s$ rolls \emph{up} and the dual universe collapses.
One may describe the evolution with  \emph{(i)} $\ha \s$ beginning in the de Sitter vacuum and  \emph{(ii)} $\ha \s$ climbing the potential towards the big-crunch, which is  \emph{(iii)} identified with the big-bang, and  \emph{(iv)} then $\ch \s$ returns down to the vacuum.  
It is important to mention that this kind of evolution is similar to the weak-coupling version of  pre-big-bang and cyclic models\footnote{The similarity to the weak-coupling regime stems from our choice of $\s < 0$.} discussed  in Sect.III of ref.\cite{turok-seiberg}; it is, however, \emph{different} from the original dilaton gravity \emph{strong-coupling} pre-big-bang scenario \cite{VezianoPrebigbangincosmo,venez-contr}.
Notice that (as it also happens in the E/E case) the SFD transformation (\ref{stils}) exchanges the ``boundary conditions'' of the scalar field, that is: the configuration \emph{(i)} is dual to \emph{(iii)}, etc. 
Finally, note that the identification of steps \emph{(i)} and \emph{(iv)}, and the endless repetition of the cycle just described is straightforward. A few applications of these results  will be discussed  in Sect.\ref{Sect.conclude}.

\section{SFD symmetric Conformal Cyclic Cosmologies} \label{Sect.CCC}

The pairs of scale factor dual FRW solutions  $(\ha a ,\ha\rho, \ha p)$ and  $(\ch a,\ch \rho, \ch p )$, joined at  their time limits ($\eta=0$ and $\eta=\pm\eta_f$),
define  pre-big-bang and cyclic cosmological models  that might  suffer of certain inconsistencies related to  discontinuities of the scale factor derivatives.\footnote{Similar to the ``graceful exit" problems of the pre-big-bang models  of dilaton gravity \cite{VezianoPrebigbangincosmo} and of the more general contracting/expanding cosmologies as well.} One expects to overcome  the eventual  ``gluing problems"  by an appropriate choice of their matter content  or else by calling for quantum or string corrections to make them smooth. The conceptually different nature of the transition region  in the expanding/expanding universes  offers however another option.

\subsection{Conformal Crossover}\label{Subsect.Crossover}

As we have shown in Sect.\ref{Sect.SFDasSymm}, the conformal crossover consists in the \emph{ identification} of the  asymptotic space-like 3D surface, representing ``de Sitter-like"  future infinity of the past aeon with the big-bang singularity  of the present aeon. In other words  we identify the  poles of $\ha a \approx 1/\eta$ with the zeros of $\ch a\approx \eta$  and  more generally $\ha a^2 \approx \eta^{1+3w}\approx 1/\ch a^2$, as required by scale factor duality (\ref{ExpanExpanDuali}).
Hence  the description of such ``conformal crossover" junction  involves a pair of metrics belonging to the  \emph{conformal (Weyl) equivalence class} of  transition metrics,
known to be the hallmark  of  the conformal cyclic cosmologies \cite{penrose, CCCcircles}. In fact the asymptotic behavior of SFD expanding/expanding universes  is nothing but a \emph{particular} realization of the Penrose's ``reciprocal hypothesis'' 
\br
\Omega \, \omega = -1,\quad\quad d \ch s^2 = \omega^2 \, ds^2 \ , \quad d\ha s^2 = \Omega^2  \, ds^2 \ ,\quad ds^2 = - d\eta^2 + d {\bf x}^2,\label{reciprocal} 
\er
that indicates how to choose one finite at $\eta=0$ (i.e. without zeros and poles) representative of this equivalence class of metrics. Since by construction  we have that $\omega$ vanishes and $\Omega$ diverges at $\eta=0$, therefore  Eq.(\ref{reciprocal}) ensures that $ds^2\propto \Omega^2 d \ch s^2 = \omega^2 d \ha s^2$ remains  finite. 
%
\begin{figure}[ht] \label{CCCSFDOmegas}
\centering
\includegraphics[scale=0.6]{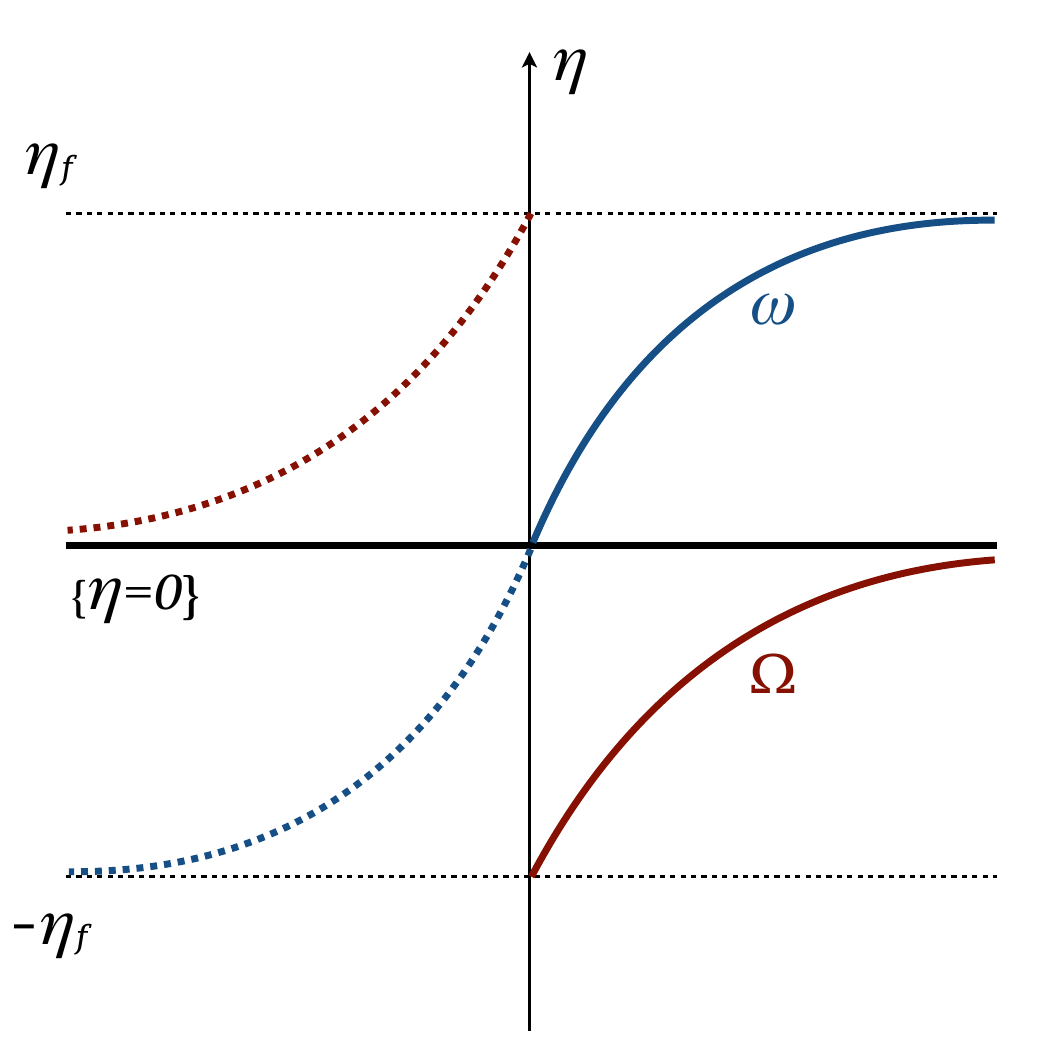} 
\caption{The conformal factors across the infinity/big-bang surface. The blue line gives $\omega(\eta)$ and the red lines give $\Omega(\eta)$; the dotted portions indicate the extensions to the opposite aeon.} 
\end{figure}
%
The negative sign in Eq.(\ref{reciprocal}) is required for $\omega$ to be a function with non-zero derivative which can be extended to the ``opposite'' aeon continuously: it simply vanishes at the big-bang and then becomes negative. Assuming that the space at the transition region is homogeneous and isotropic one can easily relate  ``reciprocal hypothesis'' (\ref{reciprocal}) to the scale factor inversions (\ref{ExpanExpanDuali}) 
\br\label{OmegaomegaCCCSFD}
 \omega(\eta) = \begin{cases} \tfrac{1}{c_0} \, \ch a(\eta)  \\ - \tfrac{1}{c_0} \, \ch a(-\eta) \\ \end{cases} \quad
{\text{and}} \quad\quad
 \Omega(\eta) = \begin{cases}  - \tfrac{1}{c_0} \, \ha a(-\eta) , & \quad  \eta > 0 \\ 
   \tfrac{1}{c_0}\, \hat a(\eta)  , & \quad \eta < 0\\ \end{cases} 
\er
as  shown   on  Fig.\ref{CCCSFDOmegas}. It gives for the conformal factors the appropriate form at their respective aeons, where they are proportional to the scale factors (e.g. $\omega^2(\eta) \sim \ch a^2(\eta)$ for $\eta > 0$). The change of sign in the arguments at the  ``opposite'' aeons (e.g. $\omega^2(\eta) \sim \ch a^2(-\eta)$ for $\eta < 0$) are necessary since the functions $\ch a(\eta)$ and $\ha a(\eta)$ are only defined, each respectively, at  the disjoint intervals $\eta \in [0 , \eta_f)$ and $\eta \in [-\eta_f, 0)$. Thus Eq.(\ref{OmegaomegaCCCSFD}) provides the most natural ``extension'' of both metrics (\ref{reciprocal}) to the whole interval $[-\eta_f, + \eta_f]$.

\subsection{Conformal Cyclic Cosmology}

The  relation between the ``reciprocal hypothesis'' (\ref{reciprocal})  and conformal time scale factor duality  (\ref{ExpanExpanDuali}) suggests that  an appropriate implementation of  SFD as a symmetry of   conformal cyclic cosmologies  might take place. 
In order to make such  a ``CCC/SFD correspondence"  more precise, we find worthwhile to  briefly stand out the main CCC characteristics.
According to the standard cosmological model the fate of our universe is to expand at an accelerated rate forever and become cold and empty. Penrose further assumes \cite{penrose}   that in the very far future (almost) all the matter will be swallowed by black holes and these, in turn, will  completely evaporate.\footnote{Due to ``supercooling" bellow  the temperature of  thermal equilibrium with Hawking radiation.} The massive particles which eventually escape falling into a black hole will, by then, have lost their rest mass due to an  ``anti-Higgs'' mechanism resulting in spontaneous enhancement of the symmetries (see \cite{penrose} for a more detailed discussion).
The universe will then be filled only with cold radiation and, therefore -- conformally (Weyl) invariant. On the other hand, the early universe, shortly after the big-bang, is also conformally invariant, filled with hot radiation, so one may conformally transform the old universe into the early one and vice-versa. The conformal identification is done, not for the beginning and the end of one same universe, but for two \emph{consecutive} eternal universes, each named an \emph{`aeon'}. 
Thus the big-bang singularity of an aeon and the conformal infinity of its predecessor are identified at one single surface $\cal X$. The metrics of each aeon, \emph{close} to this surface $\ha g_{\m\n} = \Omega \, g_{\m\n} $ and $\ch g_{\m\n} = \omega \, g_{\m\n}$
must satisfy the reciprocal hypothesis (\ref{reciprocal}).
Notice however that the pair of CCC metrics  ($\ha g_{\m\n}$ , $\ch g_{\m\n}$)  are \emph{not} in general of FRW type and therefore  the conformal factor inversion in CCC is only  an \emph{asymptotic} symmetry  that also keeps untouched the space-time causal structure.  In the transition ``crossover" region where this symmetry is valid, ``there is no one metric'', but  an equivalence classes of metrics and as a consequence, there is no notion of scales, distances or of a ``cosmic time''.  Therefore within the frameworks of the Penrose's CCC concepts for Universe evolution the massless matter, the radiation and the gravitational waves  are \emph{eternal and everpresent}, e.g.  freely transiting between the consecutive aeons and the only effect of their bypassing through the crossover surfaces $\cal{X}$ is the change in their frequencies, amplitudes, energies, etc. reflecting the conformal  principle that at $\cal{X}$'s all the distances, momenta, etc. are in fact equivalent. 

Another  distinctive feature of CCC is the so-called \emph{`suppressed rest-mass hypothesis'} (SRM). It ensures the intrinsic consistency and continuity  at $\eta=0$ for  the derivatives of the conformal factors, representing  the two (oppositely oriented ) normal vectors  $\ch N_\m=\ch\gamma\pa_\m\Omega$ and $\ha N_\m=\ha\gamma\pa_\m\omega$  to the surface  $\cal X = \{\omega = 0\}$. Written in terms of $\omega$  it reads\footnote{Notice the difference between Penrose's signature $(+---)$, the opposite of ours. Also, in Refs.\cite{penrose, CCCcircles} the authors use a different normalization of the vectors $\ch N_\m$, $\ha N_\m$, namely with  $\ch\gamma=1=\ha\gamma$, which corresponds to fixing $\ha\rho_{\Lambda}=3=\ch\rho_{\Lambda}$,  $c^2_0=1$.} 
\br
&\ha N_\m \ha N^\m= \ha\gamma^2\pa_\m \omega \, \pa^\m \omega = -1 + (2 - Q ) \omega^2 + O[\omega^3] ,\quad\quad \ch N_\m \ch N^\m|_{\cal{X}}= -1=\ha N_\m \ha N^\m|_{\cal{X}}, \nonumber\\
& \ha\gamma=\sqrt{\frac{3}{c_0^2 \ha\rho_{\Lambda}}}= \sqrt{\frac{3c_0^2 }{\ch\rho_r}}=\ch\gamma,
\quad \quad c_0^4 =\frac{\ha\rho_r}{\ch\rho_{\Lambda}}=\frac{\ch\rho_r}{\ha\rho_{\Lambda}}
\label{restmasshyp}
\er
where $Q$ is an ``universal constant'' to depend upon the matter content of both aeons and on the details of mass generation mechanism in the new aeon as well (see the Appendix A of Ref.\cite{CCCcircles}, as well as Ref.\cite{Newman_A_fundamntal_solution}). 
In fact Eq.(\ref{restmasshyp}) imposes the correct  \emph{initial/final conditions} at $\eta=0$ needed to guarantee that  the \emph{unit} normals remain time-like and their norms are close to unity up to \emph{second} order terms in the vicinity of $\cal X$. The missing linear term  is a specific requirement, forbidding that  non-relativistic matter (dust) $\rho_d \approx 1/\ha a^3$  to contribute in the neighbors of crossover $\cal{X}$; it is closely related to the initial conditions for the Yamabe equation  \cite{penrose, CCCcircles, Newman_A_fundamntal_solution},   assumed to govern the  matter induced by the conformal factor $\Omega$ in the present aeon.

The above mentioned CCC  requirements assure  the consistency of  crossover transition and the desired matter content that provides the asymptotic conformal symmetry. However they \emph{do not} exhaust  all the conditions that CCC must obey. 
As emphasized by Penrose \cite{penrose} the most important ingredient of  the Conformal Cyclic Cosmology concerns its thermodynamical (TD)  properties -- the definition of their entropy and the compatibility of  the second law with the cyclic nature of the CCC  evolution. We postpone the discussion of the  some of  thermodynamical aspects of  CCC models and the problem concerning TD consistency of their SFD symmetric versions  to Sect.\ref{Sect.TDSFD} below.

\subsection{CCC vs SFD - matter transformations} \label{Sect.matter}
 
The scale factor inversion (\ref{ExpanExpanDuali}) can be easily  recognized as a particular discrete $Z_2$ Weyl transformation $\Omega(a)$ relating the metric of  two consecutive aeons:
\br
 d\ch s^2 = \Omega^{-4} \, d \ha s^2,\quad\quad \ha a =\Omega^2(\ch a) \ch a, \quad\quad \Omega^2(\ch a)=\frac{c^2_0}{\ch a^2} , \label{weyl}
\er
as one can see from eqs.(\ref{OmegaomegaCCCSFD}). 
Notice, however, the important difference: while the SFD-symmetric CCC models  are invariant under such a restricted Weyl transformation (\ref{weyl})  within  the whole interval $[-\eta_f, + \eta_f]$,  the original CCC proposal \cite{penrose}  manifests an even more general conformal (Weyl) invariance, but in the transition region $\eta\approx 0$ only.

This observation allows to relate  the matter content of the two aeons, by adapting  the SFD methods described  in Sect.\ref{Sect.scalarsfd} to  the  case of  CCC models.
The ``field'' responsible for this mechanism is the conformal factor $\Omega$ of one aeon, when extended into the opposite one. 
For any conformal transformation of the metric, like $\ch g_{\m\n} = \Omega^{-4} \ha g_{\m\n}$, the Einstein tensor $\ch G_{\m\n} = \ch R_{\m\n} - \tfrac{1}{2} \ch R \, \ch g_{\m\n}$ transforms as 
\br
\ha G_{\m\n} = \ch G_{\m\n} + 2 \Upsilon_\m \Upsilon_\n - 2 \ch \nabla_\m \Upsilon_\n + 2 \ch g_{\m\n} \, \ch g^{\a\b} \ch\nabla_\a \Upsilon_\b + \ch g_{\m\n} \, \ch g^{\a\b} \Upsilon_\a \Upsilon_\b ,
 \label{haGchGcomUps}
\er
where $\Upsilon_\m \equiv 2 \pa_\m \log \Omega$. 
Such transformations, considered as  symmetries of  the  pre-big-bang SFD universe (and of the crossover region of  generic CCC models), have  to preserve the form of the Einstein equations at  \emph{both aeons} 
\br
\ha G_{\m\n} = \ha T_{\m\n}  \quad {\text{and}} \quad \ch G_{\m\n} = \ch T_{\m\n} ,
\er
with $\ch T_{\m\n}$ and $\ha T_{\m\n}$ appropriately related. Therefore, we must incorporate the extra terms at the right-hand-side of Eq.(\ref{haGchGcomUps}) into the stress-tensor, reading the transformation of the Einstein tensor  as an effective transformation between  the stress tensors  of the present and past aeons:
\br \label{TchThaPhi}
& \ha T_{\m\n} = \ch T_{\m\n} + t_{\m\n} ,  \;\; {\text{where}}  \\
& t_{\m\n} =\frac{4}{\phi^2} \ch\nabla_\m \phi \, \ch\nabla_\n \phi + \frac{4}{\phi} \ch\nabla_\m \ch\nabla_\n \phi -\left( \frac{4}{\phi}  \ch \square \phi - \frac{8}{\phi^2} \ch g^{\a\b} \ch\nabla_\a \phi \, \ch\nabla_\b \phi \right) \ch g_{\m\n} ,\quad {\text{with}} \quad \phi \equiv 1 / \Omega  \nonumber
\er
The tensor $t_{\m\n}$ represents the contribution to $\ha T_{\m\n}$ from the Weyl rescaling of the metric $\ch g_{\m\n}$ of its dual universe. It is then natural to consider such pairs of aeons  as  being different descriptions of  the same universe seen from two  \emph{non-equivalent conformal frames}.
Within the frameworks of Conformal Cyclic Cosmology the above results also suggest that certain gravitational degree of freedom%
\footnote{They are \emph{not}, however, ``pure gauge"  degrees of freedom, since the local conformal (Weyl) symmetry is broken outside of the transition region between the two aeons.} (encoded in $\Omega$)  is transmitted from one aeon  to the matter content of the consecutive aeon, via the peculiar double-role of the ``gravitational/matter field" $\phi$ \cite{2013arXiv1309.7248T}. 

The  SFD symmetric CCCs possess an important new feature  --- although the conformal symmetry is broken, they remain \emph{invariant} under a residual  discrete  Weyl transformation (\ref{weyl}). It   acts as an \emph{UV/IR duality symmetry} relating the large  scales behavior (at late times) of one aeon to the small scales (at earlier times) of its dual and vice-versa. Thus, given the past aeon fluid density, it allows to determine the matter content of the present aeon. In order to derive  the SFD counterpart of CCC's stress-tensor transformations we  replace $\phi = \tfrac{1}{c_0} \ch a( \eta)$ into Eqs.(\ref{TchThaPhi}).  As expected, the result of this substitution reproduces the  matter fluid SFD transformations  (\ref{sfdro}): the $00$-component is simply $\ch a^2 \,  \ch \rho = \ha a^2 \ha \rho$, while from the $ii$-components we get the pressure transformation $\ch a^2 \left(3 \ch p + \ch \rho \right) = - \ha a^2 \left( 3 \ha p + \ha \rho \right)$.

 In order to highlight the CCC features of  the SFD extensions of  modified Chaplygin gas cosmology (\ref{chaply}), we have  to  further implement   the  ``reciprocal hypothesis'' (\ref{reciprocal}) and of the asymptotic Weyl symmetry (in the neighborhood of $\cal X$)   for the  pair  $(\ha\s,\ch\s)$ of matter fields as well.  As it was shown in Sect.\ref{Sect.scalarsfd},  the corresponding partially self-dual matter Lagrangians $\ha {\cal L}( \ha\s,\ha\rho_J)$ and $\ch {\cal L}(\ch\s,\ch\rho_J)$  contain ``almost identical" potentials (\ref{poten}), with different masses $\ha m^2_{\ha\s}\neq\ch m^2_{\ch\s}$, that are related by the SFD transformations (\ref{stils}), (\ref{dualpot}) and (\ref{partmassdual}). The expanding/expanding nature of the CCC  evolution implies that we have to restrict the values of the scalar fields, say $ -\infty<\ha\s\leq 0$, thus considering only ``the expanding halves" of the potentials $\ha V$ and $\ch V$ and ``gluing" them at  $\cal X$ by conformal identification of the ``de Sitter vacua final state"  $\ha\s_{vac}= 0$  of the past aeon with the ``initial big-bang state" $\ch\s_{bb}= -\infty$ of the present aeon. Notice that the SFD mapping realized around the conformal crossover is transforming the accelerated dark energy dominated  phase $ \ha\s_{cr}\leq\ha\s\leq 0$ (i.e. small $\ha\s$  values)%
\footnote {Where $\ha\s_{cr}=\frac{1}{2\delta} \ln(1+\sqrt{2})=\ch\s_{cr}$ is determined  as a zero acceleration  value of the scalar field, $q(\s_{cr})=0$.} 
of the past aeon into the decelerated radiation dominated phase $ -\infty<\ch\s\leq\ch\s_{cr}$ (i.e. large $\ch\s$  values) of the present aeon and vice versa. The  consistent SFD description of the conformal crossover requires to further extend  the  \emph{conformal (Weyl) equivalence class} of the transition metrics to include 
the  matter fields as well.  The problem is that  the SFD transformation (\ref{stils}) of the pair of canonical scalar fields $(\ha\s,\ch\s)$ has \emph{not} the desired  Weyl form. One can use the fact that the Weyl transformations for  the scale factor are well defined in the \emph{conformal time frame} (see also Eq.(\ref{conftransfetarchitaua}))  as a guide of how to  convert the $\ch\s(\ha\s)$  transformation  into a standard Weyl form by an appropriate change of the fields variables $\s\rightarrow \Phi$, such that the infinite range of values of $\s\in(-\infty,\infty)$ is  mapped into a  finite interval for $\Phi\in[-\Phi_0,\Phi_0]$. 
 An important  hint is given by the following equivalent forms of the conformal factor $\Omega(\ch a)$ and SFD matter transformations (for $\s_0=0$):
\be
\Omega^2(\ch a)=\frac{c^2_0}{\ch a^2} =\sqrt{\frac{\ch\rho_{\Lambda}}{\ha\rho_{\Lambda}}}\left(\sinh\left(\delta\ch\s \right)\right)^{\frac{1}{\delta}},\quad \left(\frac{\ha\rho_{\Lambda}}{\ch\rho_{\Lambda}}\right)^{\delta}\tanh^2(\delta\ha\s)=\Omega^{-2\delta}\left(\frac{\ch\rho_{\Lambda}}{\ha\rho_{\Lambda}}\right)^{\delta}\tanh^2(\delta\ch\s), \label{omega}
\ee
obtained from eqs. (\ref{aetaf}) and  (\ref{stils}).  As suggested in Ref.\cite{camara}, it is then natural to introduce a new  field variable having however a \emph{non-canonical kinetic term}:
\begin{eqnarray}
&&\ha\Phi^2= \ha\phi \bar {\ha\phi}=6\left(\frac{\ha\rho_{\Lambda}}{\ch\rho_{\Lambda}}\right)^{\delta} \tanh^2(\delta\ha\s) ,\quad \ch\Phi^2= \ch\phi \bar {\ch\phi}=6\left(\frac{\ch\rho_{\Lambda}}{\ha\rho_{\Lambda}}\right)^{\delta} \tanh^2(\delta\ch\s),\nonumber\\
&& -\sqrt{6}\left(\frac{\rho_{\Lambda}}{\ha\rho_{\Lambda}}\right)^{\delta/2}\leq\ha\Phi\leq 0,\quad\quad 0\leq\ch\Phi\leq\sqrt{6}\left(\frac{\ha\rho_{\Lambda}}{\ch\rho_{\Lambda}}\right)^{\delta/2}.\label{change}
\end{eqnarray}
In fact, the substitution  (\ref{change}) transforms  the self-dual Chaplygin gas  model (\ref{poten})  into the ``gauged"  $SU(1,1)/U(1)$ K\"ahler sigma model  minimally coupled to Einstein gravity
 \begin{eqnarray} 
\ha K=-\frac{1}{2\delta^2} \ln\Big(1-\frac{\ch\rho^{\delta}_{\Lambda}}{6\ha\rho^{\delta}_{\Lambda} } \ha\phi \bar{\ha\phi}\Big) ,\quad \ha V(\ha\Phi)=\frac{\ha\rho^{2-\delta}_{\Lambda}}{9} \Big(\frac{6}{6\ha\rho^{\delta}_{\Lambda}-\ha\Phi^2 \ch\rho^{\delta}_{\Lambda}} \Big)^{\frac{1}{\delta}} (9\ha\rho^{\delta}_{\Lambda}-\ha\Phi^2 \ch\rho^{\delta}_{\Lambda}) \label{kahler}
\end{eqnarray}
with the gauge fixing of $U(1)$ symmetry realized as  $\ha\phi=\bar{\ha\phi}=\ha\Phi$, and the same for the dual fields $\ch\phi$. We have adopted the standard K\"ahler sigma model notations: 
\be
\ha {\cal L}_{matter}(\ha\phi ,\bar{\ha\phi})= g^{\ha\phi \bar {\ha\phi}} \ha \nabla_{\mu} \ha\phi \ha\nabla^{\mu} \bar {\ha\phi} - \ha V(\ha\Phi) ,\quad \quad g^{\ha\phi \bar {\ha\phi}}= \partial_{\ha\phi} \partial_{\bar {\ha\phi}} \ha K,\label{kalmetric}
\ee
where, due to the specific choice of the K\"ahler  potential $\ha K(\ha\phi ,\bar{\ha\phi})$ in Eq.(\ref{kahler}), the  resulting K\"ahler  metric  $g^{\phi \bar {\phi}}$ has a constant curvature $R_{K} =-4 \delta^2$.
Within the framework of the  above ``K\"ahler representation" of  CCC model,  the new matter fields SFD transformation takes the proper Weyl form we were seeking:
\be
\ha\Phi =\Omega^{-2\delta} \ch\Phi, \quad\quad  \ha a =\Omega^2(\ch a) \ch a, \quad\quad \Omega^2(\ch\Phi)=\frac{c^2_0}{\ch a^2} =\Big(\frac{\ch\Phi^2 \ch\rho^{\delta}_{\Lambda}} {6\ch\rho^{\delta}_{\Lambda}-\ch\Phi^2 \ha\rho^{\delta}_{\Lambda}}\Big)^{\frac{1}{2\delta}}.\label{weylphi}
\ee
Let us  mention one ``unexpected SUSY feature"  of this  K\"ahler sigma model realization of  SFD-symmetric CCC  (\ref{kalmetric}) discovered in ref.\cite{camara}:  it turns out to be  identical with the bosonic  part of the  ${\cal N} =1$ supergravity with one chiral matter supermultiplet \cite{ferrara,kalosh} and a very special  form  of its matter superpotential  $|\ha {\cal Z}|= \sqrt{\frac{\ha\rho_{\Lambda}}{3}} e^{\delta \ha K} =\sqrt{\frac{2}{3} \ha\rho}$, that gives rise  to the partially self-dual  potential (\ref{kahler}).

The above discussion  of the CCC properties of  the pre-big-bang  extension of   modified Chaplygin gas model  (\ref{chaply})  has demonstrated the advantages of the SFD methods in the selection of the matter fields self-interactions as well as in the description of their  behavior both around  the conformal crossover and  far from it. We have also accumulated  several  evidences that its  K\"ahler form  appears to be a promising candidate for SFD-symmetric CCC with physically relevant  matter content  of a certain supergravity origin.  It remains, however, to verify whether it  also satisfies all the other CCC conditions \cite{penrose}.

\subsection{CCC vs SFD - suppressed rest-mass conditions} \label{SectCCCSFDSRM}

The SFD counterpart of Penrose's SRM hypothesis can be obtained  by replacing  the SFD form (\ref{OmegaomegaCCCSFD}) of the conformal factor $\omega$  into Eq.(\ref{restmasshyp}), which now reads:
\bsub\label{SRMHFRW}\br
&&  \ch\gamma^2\pa_\m\omega \, \pa^\m\omega = -1 + \tfrac{(2-Q)}{c_0^2} \ch a^2(\eta) + O[\ch a^3(\eta)], \quad\text{for} \quad \eta > 0 , \label{SRMHFRWa}\\
&& \ha\gamma^2 \pa_\m\omega \, \pa^\m\omega = -1 + c_0^2 (2 - Q) \ha a^{-2}(\eta) + O[\ha a^{-3}(\eta)], \quad \text{for} \quad \eta < 0 .
\er\esub
On the other hand, by using once again  Eq.(\ref{OmegaomegaCCCSFD}) for $\omega$, we can directly calculate the norms of the vectors $\ch N_\m=\ch\gamma\pa_\m\Omega$ and $\ha N_\m=\ha\gamma\pa_\m\omega$,
\bsub\label{paopaoFreqmai0}\br
 \ch\gamma^2\pa_\m\omega \, \pa^\m\omega &=& - \tfrac{3}{\ch\rho_r} (\ch a ' (\eta) )^2 =  - \tfrac{1}{\ch\rho_r} \, \ch a^4 \, \ch \rho, \quad\text{for} \quad \eta > 0, \\
 \ha\gamma^2\pa_\m\omega \, \pa^\m\omega &=& - \tfrac{3}{\ha\rho_{\Lambda}} \left( \tfrac{\ha a'(\eta)}{\ha a^2(\eta)} \right)^2 = - \tfrac{1}{\ha\rho_{\Lambda}}   \, \ha \rho, \quad\text{for}\quad \eta < 0. 
\er\esub
At the last step of these equalities we have used the Friedmann equations. Note that \emph{both} the norms of $\ch N_\m$ and $\ha N_\m$ turn out to be proportional to the density of the first aeon $\ha\rho=\tfrac{\ch a ^4}{c_0^4}\ch\rho$. 
The identification of the r.h.s. of Eqs.(\ref{SRMHFRW}) and (\ref{paopaoFreqmai0}) furnishes the explicit SFD form  of  the original SRM condition  (\ref{restmasshyp}).

When the expansion (\ref{SRMHFRW}) contains only integer power of the scale factor, we may include the terms up to fourth order,
\br
 \ch\gamma^2\pa_\m\omega \, \pa^\m\omega = -1 + \tfrac{(2-Q)}{c_0^2} \ch a^2 - \tfrac{\ch Q_3}{c_0^2} \ch a^3 - \tfrac{\ch Q_4}{c_0^2} \ch a^4  + \cdots \quad\text{for} \quad \eta > 0 ,	\label{expNintegpower}
 \er
 and accordingly for $\eta < 0$, so the implementation of the SRM condition on SFD models may be written as
\bsub\label{rhoexp}\br
&&  \ch \rho  =\frac{\ch\rho_r}{\ch a^4}  + \frac{\ch\rho_{str}}{\ch a^2} + \frac{\ch\rho_{dw}}{ \ch a} + \ch\rho_{\Lambda}+ \cdots  , \quad\text{for}\quad \ch a \to 0 , \label{rhoexpa} \\
&&  \ha \rho = \ha\rho_{\Lambda}  + \frac{\ha\rho_{str}}{\ha a^2} + \frac{\ha\rho_d}{\ha a^3} +  \frac{\ha\rho_r}{\ha a^4} + \cdots ,\quad\text{for} \quad \ha a \to \infty 	. \label{rhoexpb}
\er\esub
Thus, stoping at fourth order in the expansion (\ref{expNintegpower}),  the universe is filled with a composition of perfect fluids, and the original SRM requirement that the  term linear in $\omega$ must be excluded from the expansion in (\ref{SRMHFRWa}) is, as expected, equivalent to the absence of a dust component in $\ch \rho$ on the beginning of the the post-big-bang phase, i.e. $\ch \rho_d = 0$. Meanwhile, the requirement of normalization of $\ch N_\m$ and $\ha N_\m$ imply that the dominating components, near $\cal X$, at the corresponding aeons are radiation ($\ch \rho_r / \ch a^4$) and a cosmological constant $\ha \rho_\La$. The second order term in (\ref{SRMHFRWa}) turns out to be related to the presence in both aeons of a string gas component, with relative densities $\ch \rho_{str} = \frac{\ch \rho_r}{c_0^2} (Q - 2)$ and $\ha \rho_{str} = c_0^2 \ha \rho_\La (Q - 2)$.%
\footnote{Therefore the condition $Q = 2$ obtained in Ref.\cite{Newman_A_fundamntal_solution} is equivalent to the absence of a string gas. The result that the aeons are filled by perfect fluids as a consequence of an expansion around the crossover has also been found in \cite{poloneses}.} 
Similarly, $\ch Q_3$ is related to a gas of domain walls ($\ch\rho_{dw}$), and $\ch Q_4$ to a cosmological constant $\ch \rho_\La$ in the post-big-bang phase.
We have
\br
& \ch \rho_{dw} = \tfrac{1}{c_0^3} \ch Q_3 \ch \rho_r \; , \quad \ch \rho_\La = \tfrac{1}{c_0^4} \ch Q_4 \ch \rho_r \; ; \quad \quad \ha \rho_d = c_0^3 \ha Q_3 \ha \rho_\La \; , \quad \ha \rho_r = c_0^4 \ha Q_4 \ha \rho_\La .	\label{Qexpan}
\er
The coefficient $Q$ in Eq.(\ref{restmasshyp}) is assumed by Penrose \cite{CCCcircles} to be ``universal'' in the sense that it is the same in avery aeon. If we assume this of the other coefficients, so that $\ch Q_3 = \ha Q_3$ and $\ch Q_4 = \ha Q_4$, then Eqs.(\ref{Qexpan}) fixes relations between ratios of densities in both aeons, viz. 
\br
\ha \rho_{str} / \ch \rho_{str} = c_0^4 ( \ha \rho_\La / \ch \rho_r) \; , \quad \ha \rho_d / \ch \rho_{dw} = c_0^6 ( \ha \rho_\La / \ch \rho_r) \; ,  \quad \ha \rho_r / \ch \rho_\La = c_0^8 ( \ha \rho_\La / \ch \rho_r).	\label{SRMrhoratios}
\er 
These are the same relations imposed by SFD with, also, $\ha \rho_r / \ch \rho_\La = c_0^4$ (cf. Eq.(\ref{LaCDMdaul})). We may then turn the argument around and conclude that SFD fixes $\ch Q_3 = \ha Q_3$ and $\ch Q_4 = \ha Q_4$.

The allowed presence of the dust component $\ha \rho_d / \ha a^3$ in the pre-big-bang phase stems from the fact that we are considering its ``future asymptotic'', and the SRM condition forbids the presence of dust in the ``past asymptotic'', i.e. near the big-bang. But we have been considering only one pair of dual aeons. We now turn to the implementation of the SRM condition on the cyclic extension composed of an infinite chain of aeons described in Sect.\ref{CyclciSFDextensions}. Then, the surface $\ha a = 0$ is another crossover, in the vicinity of which $\ha \rho$ must satisfy an expansion of the same form of (\ref{rhoexpa}), i.e. we must have $\ha \rho_d = 0$ and this reflects on the following aeon, since Eq.(\ref{SRMrhoratios}) then forces $\ch \rho_{dw} = 0$. Therefore the SRM condition, imposed on a chain of SFD aeons with perfect fluids, forbids \emph{both} dust \emph{and} its dual, a gas of domain walls.

This prohibition of even a late-time dust component moves us to consider models which are more elaborate than the simple composition of independent perfect fluids. The most relevant here is the modified Chaplygin gas density (\ref{chaply}), whose asymptotic limit near $\cal X$,
\be
\ch\rho=\ch a^{-4} (\ch\rho^{\delta} _r+\ch\rho^{\delta}_{\Lambda} \ch a ^{4\delta})^{1/\delta}= \ch\rho_r \ch a^{-4} + \tfrac{\ch\rho^{\delta}_{\Lambda}}{\delta\ch\rho^{\delta}_r }\ch a^{-4(1-\delta)} + O[\ch a^{-4(1-2\delta)}] \quad {\text{for}} \quad \ch a \to 0	\label{ChapasymX}
\ee
is  indeed ``SRM consistent" if  $\delta\in[1/2,1]$.  
If we also assume that the expansion is in  integer powers of $\ch a$,  then  the SRM condition selects  three values of  $\delta= \frac{1}{2}, \frac{3}{4}, 1$.  The case with $\delta = \frac{3}{4}$ represents a particularly interesting SFD-symmetric CCC model,  that   manifests many of the properties of the  Chaplygin-like \emph{quintessence} models \cite{chap-modif1, chap-mod2}:  at early times ($\ch a \approx 0$) it is approximated by radiation and domain walls $\ch \rho \approx \ch\rho_{r}/\ch a^4+\ch\rho_{dw}/\ch a$, while at late times ($\ch a \approx \infty$)  it gets a contribution from the cosmological constant and dust, $\ch\rho \approx \ch\rho_{\Lambda} +\ch \rho_{d}/\ch a^3$. In other words, it is an example of the desired cosmological model which has a late-time dust-like behavior while still respecting the SRM condition. Note that since every  aeon in a SFD chain is filled with a Chaplygin gas with the same $\delta$ (but possibly different values of $\rho_r$ and $\rho_\La$), the consistency of $\ch \rho$ in Eq.(\ref{ChapasymX}) above is sufficient to assure the consistency on every crossover of the chain. Note also that both the Chaplygin gas and  the consistent composition of perfect fluids discussed above turn out to be partially self-dual. 

When we consider the scalar field counterpart of the Chaplygin gas, the condition $\delta= \frac{1}{2}, \frac{3}{4}, 1$ fixes the curvature of the K\"ahler metric to be  $R_K=-1,-\frac{9}{4},-4$.
But the most important consequence of the SRM conditions is that they provide consistent initial and final conditions (on $\cal X$)  not only for the scale factor $\ha a$ and its dual $\ch a$, but for the pairs fields $(\ha\s,\ch\s)$ and $(\ha\Phi, \ch\Phi)$ as well --- it is enough to  take into account  Eqs.(\ref{scaltr}), (\ref{aetaf}) and (\ref{change}). In other  words they guarantee the ``crossover consistency" for the solutions of the matter fields equations of the considered \emph{partially self-dual massive} $SU(1,1)/U(1)$ K\"ahler sigma model (\ref{kalmetric}).

\subsection{CCC and SFD - comparison}

Our investigation of the consequences of the implementation of scale factor duality to Penrose's CCC models has revealed certain similarities and diferences, in comparison with SFD-symmetric pre-big-bang and cyclic cosmologies. Their main common feature is the conformal crossover. In fact,  both  represent two different (but interrelated) implementations of the same symmetry principle of \emph{restricted and/or asymptotic} conformal (Weyl) invariance. In  SFD  models, the scale factor inversion is an exact symmetry between the \emph{whole} evolution of the universe background and its dual.  It is, however, more restrictive then the  CCC's asymptotic symmetries  --- SFD is  a  symmetry of the Friedmann equations only,  and thus not valid when fluctuations are considered. Furthermore, it is not a ``full'' conformal transformation, but rather a particular  discrete  $Z_2$  subgroup of it.  One consequence of these diferences is that   SFD models   satisfy  rather rigid  restrictions on the matter contents of the two aeons, while in the CCC models  the  ``suppressed rest-mass" condition, although  compatible with scale factor duality, turns out to be much  less  restrictive  since it acts only in the vicinity of the crossover. Therefore, one may regard

\vspace{0.2cm}

\noindent
\emph{SFD  as a ``broken extension'' of CCC: ``broken'' because the scale  factor inversion is a particular (discrete) Weyl transformation,  and ``extended'' because it isn't valid at a small vicinity of the crossover only.}

\vspace{0.2cm}

The main result  presented in this section is that non-trivial SFD-symmetric CCC models \emph{do  exist}. In other words, the CCC requirements  may be implemented within the frameworks of the cyclic SFD  expansion/expansion models by self-dual SFD cyclic cosmologies with manifest double $Z_2\times Z_2$ Weyl symmetry (\ref{weylphi}) (see also Eqs.(\ref{sfdca}) and (\ref{doublesfd})). They provide explicit examples of  consistent CCC models, whose matter content is given by certain gauged K\"ahler sigma models (\ref{kahler}), (\ref{kalmetric}).

Within the considered  ``SFD/CCC  context", we can also have  SFD symmetric expanding/expanding cosmologies that \emph{are not exactly of CCC-type}, in which the  suppressed rest-mass hypothesis is not satisfied but the continuity of the derivatives of the conformal factors is respected. In other words, the normal vectors  $\ch N_\m=\ch\gamma\pa_\m\Omega$ and $\ha N_\m=\ha\gamma\pa_\m\omega$  are continuous on the crossover surfaces $\cal X = \{\omega = 0\}$, but we may include the  term linear in $\omega$ in Eqs.(\ref{restmasshyp}) and (\ref{paopaoFreqmai0}).  This ``weak CCC condition'' is compatible, for example, with the $\Lambda$CDM model
\be  
\rho_{{\rm{\Lambda CDM}}} =\rho_{\Lambda}+\frac{\rho_{r}}{a^4}+\frac{\rho_{d}}{a^3},
 \label{lcdm}
\ee
which we note that, however, produces a sequence of aeons alternating with universes which contain domain walls instead of dust.

 \section{On the thermodynamics of SFD symmetric CCC models}\label{Sect.TDSFD}

The interest in studying the thermodynamics of  ``unconventional  cosmologies" such as the pre-big-bang versions of dilaton gravity \cite{VezianoPrebigbangincosmo,venez2} and the CCC  models \cite{penrose}  has to do with the attempts to explain the low entropy of the early universe and/or  to reach an alternative description of its inflationary phase.  Despite the complexity of the thermodynamical phenomena 
that occur during the evolution of the observable universe and the lack of complete understanding of the gravitational contributions to the universe entropy, certain model independent estimations of  the asymptotic initial and finite values of appropriately defined generalized entropy  are available \cite{hawking,wald} (see also refs.\cite{bousso2016generalized, bousso2015new}). Their implementation  to the case of considered  SFD symmetric pre-big-bang and cyclic cosmologies is \emph{not}  straightforward and it turns out to involve several additional requirements and restrictions.  The desired thermodynamic features, expected to takes place in the original CCC scenario \cite{penrose},  are known to be quite far from the  relatively simple adiabatic thermodynamics of the considered partially self-dual SFD symmetric cyclic models. Nevertheless,  the established  SFD/CCC relations  allows us to gain some important insights on the universal asymptotic (near the crossover) thermodynamical features of the proper CCC models as well.

\subsection{SFD Thermodynamics}

The thermal history of each ``SFD aeon", filled by a barotropic fluid,%
\footnote{We assume here that $\rho(T)$ and $p(T)$ are functions of the temperature, and that the entropy $S(V,T)$ of the physical volume $V=V_0 a^3$, where $V_0$ is a given constant comoving volume, is an extensive quantity with density $s$.}  
is known to be an adiabatic  process (see, e.g. \cite{weinberg}) characterized by
\begin{eqnarray}
s=\frac{1}{T}(p+\rho)=\frac{S}{V} ,\quad\quad E=ST-pV, \quad\quad ds=\frac{d\rho}{T},\quad\quad \frac{d\rho}{dT}=\frac{1}{T}\left(p+\rho\right)\frac{d\rho}{dp}, \label{TDcons}
\end{eqnarray}
derived from the first law of thermodynamics%
\footnote{Which is known to be equivalent to the Friedmann equations (\ref{frw}), and vice-versa \cite{jacobson}.} 
$dE=TdS - pdV$, where  $E=\rho V$ denotes the internal energy within the physical volume $V=a^3 V_0$. For each  fluid with  a given EoS $p=w(\rho)\rho$, the above Eqs.(\ref{TDcons}) allow to deduce   the thermal evolution  of  $\rho(T)$, $s(T)$  and  $a(T)$.
One of the most important  thermodynamical features of such fluids is that the $2^{\rm{nd}}$ TD law, $dS/dt \geq 0$, is trivially satisfied, since the comoving entropy density $\mathcal{S}$ and the entropy of the physical volume $V$ remain \emph{constant} during the universe evolution: 
\begin{eqnarray}
sa^3=\mathcal{S} = const,\quad\quad  S=\mathcal{S} V_0=const.\label{constentr}
\end{eqnarray}

The SFD transformations of the thermodynamical quantities  of the consecutive aeons, say $\ha S=\ha S(\ch S)$, etc., may be derived by taking into  account Eqs.(\ref{sfdciclic}) together with the $E$, $S$ and $T$ relation (\ref{TDcons}), which allows to exclude the pressure $p$, so that
\be
\frac{E_j}{a_j}=\frac{E_{j+1}}{a_{j+1}},\quad\quad \frac{1}{a_j}\left(S_jT_j - \frac{2}{3} E_j\right)= - \frac{1}{a_{j+1}}\left(S_{j+1}T_{j+1} - \frac{2}{3} E_{j+1}\right).\label{tdsfd}
\ee 

We next consider the modified Chaplygin gas models (\ref{chaply}), whose thermodynamic evolution may be written as
\begin{eqnarray}
\rho\left(1-\frac{\rho^{\delta}_{\Lambda}}{\rho^{\delta}}\right)^{4-\frac{3}{\delta}}= 4\,\sigma \, T^4,\quad\quad \mathcal{S}\equiv sa^3=\frac{4}{3}\left(4\sigma \rho_r^3\right)^{1/4},\quad \infty > \rho\geq\rho_{\Lambda},
\label{r_T}
\end{eqnarray}
obtained from the last of the Eqs.(\ref{TDcons}). The  constant of integration $4\sigma$, where  $\s=\tfrac{\pi^2 K^4_B}{60\hbar^3}$ is the Stephan-Boltzmann constant ($c=1$ is assumed in all the formulae), has been chosen  in such a way that for pure radiation, i.e.  for $\rho_{\Lambda}\rightarrow0$, Eq.(\ref{r_T}) becomes the  Stephan-Boltzmann law,  $\rho=4\, \s \,T^4$. Notice that the above expression for $T(\rho)$ leads to  three qualitatively different thermal histories \cite{sotkov2016thermodynamics}: 

\bul For $0<\delta <\frac{3}{4}$, the temperature $T(a)$ is \emph{not a monotonic} function, and diverges at both limits $\rho\rightarrow\rho_{\Lambda}$  and $\rho\rightarrow\infty$. Thus the initial big-bang stage \emph{and} the final (nearly de Sitter) stage are equally hot.  The behavior of  such models is in \emph{disagreement}  with  the expected   cooling down of the expanding universe along with the (monotonic) decrease of its curvature. The origin of this rather \emph{unphysical} behavior turns out to be related to their  thermodynamical instability \cite{sotkov2016thermodynamics}, namely  the sound velocity square $v^2_s=\tfrac{dp}{d\rho}$ for the models with $\delta<\tfrac{3}{4}$, becomes \emph{negative} for a certain range of values of the fluid density \footnote{For a proof of this statement, together with  the detailed description of TD features of the considered  SFD symmetric cosmologies, see our recent preprint \cite{sotkov2016thermodynamics}.}.

%
%

\bul  For all the models with $\delta >\frac{3}{4}$, the temperature  is monotonically decreasing for  $T\in (\infty,0) $.

\bul One of the preferred CCC fluids, with $\delta =3/4$ (see end of Sect.\ref{SectCCCSFDSRM}), gives rise to a rather interesting thermal evolution with a \emph{non zero} finite temperature $T_{\Lambda}=\left(\tfrac{\rho_{\Lambda}}{4\s}\right)^{1/4}=T_*$, and with a Stephan-Boltzmann law  almost identical  to the radiation one.

Hence the modified Chapliging gas models (\ref{chaply}) are thermodynamically consistent for $\delta$ within the interval $3/4\leq\delta\leq 1$ only. Let us remember that the upper limit is imposed in order  to avoid certain  ``unphysical" behavior of the Ricci curvature $ R=2 \rho^{\delta}_{\Lambda} \left( \rho \right)^{1-\delta}$, which for $\delta>1$ vanishes at the (still singular) big-bang, and then increases up to a finite value $R_{\Lambda}=2 \rho_{\Lambda}$.

 The advantage of SFD symmetric models is that once we know the matter content and its thermodynamical characteristics at a given aeon (say, at arbitrary instant $-\eta_*$), then the SFD transformations uniquely determine their values at all the other aeons. 
The temperature and the entropy of a ``past aeon" $\cal {\ha A}$ filled by a modified Chaplygin gas with $\delta\in[3/4, 1]$ is given by Eqs.(\ref{r_T}). By construction,  the ``present aeon''  $\cal {\ch A}$ has the same matter content (\ref{chaply}), but with different relative densities $\ha\rho_r\neq\ch\rho_r$ and $\ha\rho_{\Lambda}\neq\ch\rho_{\Lambda}$, while the physical volume entropies are  related by the SFD transformations as $\ch S= \left(\frac{\ch\rho_r}{\ha\rho_r}\right)^{\frac{3}{4}} \ha S $.
Although the direct relation between temperatures, $\ch T=\ch T(\ha T)$, has a rather complicated form, in the limiting cases $\delta=3/4$ and $\delta=1$ it simplifies to:
\br
&\delta=\frac{3}{4} :\quad \ch {\cal T}^3=\frac{\ha {\cal T}^3}{\ha {\cal T^3}-1};\quad\quad 
\delta=1: \quad  \ch {\cal T}=\left(\frac{\ha\rho_{\Lambda}}{\ch\rho_{\Lambda}} \right)^{\frac{1}{2}} \frac{1}{\ha{\cal T}}; \quad \cal  {\ha T} =\frac{\ha T}{\ha T_*},\quad \ha T_*=\left(\ha\rho_{\Lambda}\tfrac{c}{4\s}\right)^{1/4}\label{temp}
\er
The above transformations demonstrate that under certain additional restrictions on the EoS of considered Chapligin-like fluids, i.e. for $3/4\leq\delta\leq 1$ only, the UV/IR features of the SFD symmetry takes place. Namely, the late time low temperatures characterizing the accelerated phase of $\cal {\ch A}$ are mapped to the early time (nearly big-bang) high temperatures in the decelerated  phase of $\cal {\ha A}$. 
In the case of self-dual models, despite the changes of the energy and the temperature  in the vicinity of the crossover, the entropy of the considered physical volumes remains unchanged, i.e.  we have $\ha S=\ch S$  due to $\ha\rho_r=\ch\rho_r$. We should mention that the TD's of a given physical volume is ``causally consistent" when considered far enough from the crossover. It becomes however  problematic  in the vicinity of  the crossover (when it is approached  from the final de Sitter  region $\ch a\rightarrow\infty$ of the past aeon), since now a part of the physical volume becomes  \emph{unobservable}  ---  causally disconnected due to the fact that its size overpasses the event horizon. 
A more consistent  description of  the thermodynamics of the observable universe can be reached by considering the entropy of the matter within the event or apparent horizons.

\subsection{SFD for Horizons Thermodynamics}

The apparent horizon is a marginally anti-trapped  spherical surface centered at the observer position, its physical radius  $\Ups_A= 1 / | H |$ (for $K=0$), coincides with the Hubble radius where the velocity of space-time expansion becomes grater than the speed of light. Because of its (local) causal nature, one may  assign a  kind of  Bekenstein-Hawking thermodynamics to apparent horizons, with entropy and temperature%
 \footnote{Recall that we use units in which $G = 1 / 8 \pi$.}  \cite{caikim,cai-akbar} 
\br
 T_A = \frac{1}{2\pi \Ups_A}\; , \quad\quad S_A = \frac{1}{4G} (4 \pi \Ups_A^2) = \frac{8 \pi^2}{H^2}, \label{CaiKim}
 \er
such that the  Clausius relation $dQ = T_A dS_A $ indeed holds as a consequence of the Friedmann equations (\ref{frw}) and vice-versa \cite{caikim}.  Due to the relations $H =\sqrt{\rho/3}=1/\Ups_A$,  the SFD transformations of the  apparent horizon TD quantities can be easily obtained from Eqs.(\ref{ExpanExpanDuali}):
\br
& \ha\Ups_A / \ha a = \ch\Ups_A / \ch a \quad ({\text{that is}} \quad  \ha  r_A = \ch r_A) , \quad {\text{and}} \quad \ha a \ha T_A = \ch a \ch T_A ,\quad \ha S_A / \ha a^2 = \ch S_A / \ch a^2 , \label{UpsUpsa}
\er
where $r_A$ is the apparent horizon's comoving radius.

In the case of partially self-dual SFD symmetric CCC models based on modified Chaplygin gas (\ref{chaply}), the entropy dependence of the scale factor reads:
$a^{-4\delta}=\frac{\rho^{\delta}_{\Lambda}}{\rho^{\delta}_r} \left(\frac{S^{\delta}_{\La}}{S^{\delta}_A}-1\right)$. 
Then, given the past aeon apparent horizon entropy $\ha S_A$, the  SFD transformation (\ref{UpsUpsa}) determines the  entropy $\ch S_A$ at the present aeon as
\be
\ch S_A=\frac{\ch S_{\La}}{\ha S_{\La}}\left(\ha S^{\delta}_{\La}- \ha S^{\delta}_A \right)^{\tfrac{1}{\delta}},\quad\quad  S_{\La} \equiv \frac{24\pi^2}{\rho_{\Lambda}},\label{horts}
\ee
with $S_{\La}$  denoting the maximal value of the apparent horizon entropy in the asymptotic future, defined as one forth of the event horizon area at the crossover. It coincides in this limit with the  event  horizon entropy (and indeed with the actual space-time entropy) of de Sitter space.

The entropy of the observed universe includes, of course, the contribution of the entropy $S_f$ of the matter within the apparent horizon. So we have the total (generalized) entropy 
\br
&S_t = S_f+S_A=s \times V_A + 2\pi A_A , \label{totentropy} \\
& V_A = \tfrac{4}{3} \pi H^{-3},\quad\quad A_A = 4 \pi H^{-2},
\quad \quad S_f=\mathcal{S}\frac{4\pi}{3}r_A^3=\frac{4\pi}{3}\frac{\mathcal{S}}{(aH)^3},\quad\quad  r_A = 1/ (aH), \nonumber
\er
where $V_A $  e $A_A $ are the physical volume and area of the apparent horizon. 
The transformation of this total entropy between aeons is non-homogeneous and given by
 \br
\ch S_t = \left(\frac{\ha\rho_{\Lambda}}{\ch\rho_{\Lambda}}\right)^{\frac{3}{4}} \ha S_f + \left(\frac{\ha\rho_{\Lambda}}{\ch\rho_{\Lambda}}\right) \ha S_A \left(\frac{\ha S^{\delta}_{\La}}{\ha S^{\delta}_A} - 1 \right)^{\frac{1}{\delta}} , 	\label{totepartnonGB}
\er
which in the self-dual case, when $\ha\rho_{\Lambda}=\ch\rho_{\Lambda}$, simplifies to
\br
\ch S_t=\ha S_t + \ha S_A \Big[\left(\frac{\ha S^{\delta}_{\La}}{\ha S^{\delta}_A} - 1\right)^{\tfrac{1}{\delta}} -1\Big]  .
\label{sfdtotentrnonGB}
\er
The proof is a straightforward consequence of  the $S_A$  transformations (\ref{horts}) and of the fluid entropy SFD law $\ha S_f= \left(\frac{\ch\rho_{\Lambda}}{\ha\rho_{\Lambda}}\right)^{\frac{3}{4}} \ch S_f $.

Differently from the constant entropy of an arbitrary physical volume (see Eq.({\ref{constentr})), 
the matter fluid  entropy $S_f$ is now time dependent  and it turns out to satisfy the $2^{\rm{ nd}}$  law
in a decelerated matter/radiation phase only, i.e. when $q(a)=\frac{1}{2}\left(1+3\frac{p}{\rho}\right)\geq 0$, since
\be
\dot S_f=\frac{2\pi}{3}\frac{\mathcal{S} a}{(aH)^4} (\rho+3p)\geq 0.\label{fluidseclaw}
\ee
The origin of the violation of the $2^{\rm{ nd}}$  law for $S_f$ during the accelerated phase of the universe (i.e for  $q < 0$)  is related  to the decreasing  of the comoving volume $V_A$ of the apparent horizon in this case.

The  problem we are obliged to face up  here is about  the conditions  that might ensure the validity of the generalized $2^{\rm{ nd}}$ law, $\dot S_t\geq 0$, during the evolution at a given aeon.  For the modified Chaplygin gas (\ref{chaply}), we have 
 \begin{eqnarray}
\frac{dS_t}{da}=\frac{96 \pi^2 \rho^{\delta}_r}{a^{4\delta+1}}\frac{1}{\left(\rho^{\delta}_{\Lambda}+\rho^{\delta}_r a^{-4\delta}\right)^{\frac{1}{\delta}+1}} \left[1+\frac{\sqrt{3} \mathcal{S}}{8\pi \rho^{\delta}_r a^3}\frac{\left(\rho^{\delta}_r - \rho^{\delta}_{\Lambda} a^{4\delta}\right)}{\left(\rho^{\delta}_{\Lambda}+\rho^{\delta}_r a^{-4\delta}\right)^{\frac{1}{2\delta}}}\right]\geq 0, \label{col}
\end{eqnarray}
where we have used $H=\sqrt{\frac{\rho}{3}}$  with $\rho$ given by Eq.(\ref{chaply}). We must consider the validity of the above inequality only within the physical interval where $3/4\leq\delta\leq 1$. The result may be summarized  in the following statement: 

\vspace{0.2cm}
The $2^{\rm{ nd}}$  law holds during the entire  evolution of the observable universe, i.e. for all $\eta\in[0,\eta_f]$,  \emph{only  for the $\delta=3/4$ model}, and provided  its vacuum density  satisfies a distinct  \emph{upper bound}
\begin{eqnarray}
\rho_{\Lambda}^{1/4}\leq\frac{2\pi \sqrt{3}}{\left(4\sigma\right)^{1/4}}.\label{3/4}
\end{eqnarray}
($c=1=\hbar$; $\s$ the Steffan-Boltzmann constant.)
 This restriction on the maximal allowed value of the cosmological constant introduces  a \emph{lower bound}, $S_{\La}\geq\frac{2\sigma}{3\pi^2}$, on the maximal entropy of the universe filled with the $\delta=3/4$ modified Chaplygin gas.
 
 \vspace{0.2cm}
 Its proof is presented in Appendix \ref{Sect.delta}. This result establishes the TD consistency of a distinct cosmological model (\ref{chaply}) within the framework of one particular  thermodynamical description of the observable universe involving  apparent horizons \cite{caikim,cai-akbar}. It also illustrates the difficulties inherent to the definition of a generalized entropy in cosmology. When the universe enters an accelerated phase, the comoving radius of the apparent horizon shrinks, as noted in Eq.(\ref{fluidseclaw}), and in fact it vanishes as one asymptotes towards de Sitter space-time. Therefore the total (adiabatic) entropy of the fluid inside it naturally decreases, as the sphere inside this horizon becomes empty. If one simply adds this decreasing entropy of the fluid with the Bekenstein-Hawking entropy of the horizon itself, the resulting ``total entropy'' $S_t$ will only obey the $2^\nd$ Law for very specific cases in which the increasing (physical) area dominates the decreasing (comoving) volume. Eq.(\ref{col}) shows that, although relatively ``rare'', such universes \emph{do exist}. It is more natural however to seek for another definition of the generalized entropy that ensure the validity of the $2^\nd$ Law for  the  observable universe  filled by physically relevant fluid densities, say for the $\Lambda$CDM model at least. A different approach to  the total entropy, based on a new area-theorem \cite{bousso2015new}, was recently proposed in  Ref.\cite{bousso2016generalized}. Their assertion is that the generalized entropy --- matter entropy inside a volume, plus one quarter of the corresponding area --- always increases monotonically along a specifically constructed hypersurface called the holographic  Q-screen, which  (generally) differs slightly from the apparent horizon.

\subsection{Gauss-Bonnet extension and Crossover Entropy}	\label{Sybsect.GBcross}

Once we  choose to work with the generalized entropy (\ref{totentropy}), the problem arises of whether one can extend the above results to the cyclic SFD symmetric CCC models  ---  what are the eventual additional conditions  to be imposed on the behavior  of the entropy $S_t$  in the vicinity of the crossover? Should we assign a certain amount of entropy to each crossover surface $\cal X_j$? 
Must the generalized entropy  also increase from one aeon to the next one during the cyclic evolution of the universe?

The first problem to be solved, however, concerns the
\emph{zero value} of the total entropy at  the big-bang in a radiation dominated early universe: for $a \to 0$,
\br
S_t \approx \frac{4\pi \sqrt{3}}{{ \rho_r^3}}  \mathcal{S} \;  a^3 + \frac{24\pi^2}{ \rho_r} a^2 \to  0 .
\er
 Such an ``unphysical" zero entropy initial state of the universe evolution is due to the fact that the  radius of the apparent horizon vanishes at the big-bang. Therefore, on the vicinity of the crossover, its area  $A_A = 4 \pi H^{-2}$ becomes smaller than the Planck area,%
 \footnote{In the units used in this paper, with $8\pi G=1$, we have $A_{pl}=2\pi$, i.e. $l^2_{pl}=1/2$.}
i.e. $A_A < A_{pl}=4\pi l^2_{pl}= 16 \pi^2 G$, which is beyond  of the scale of validity of \emph{classical} Einstein gravity.
One solution to this problem  consists in the modification of the apparent horizon's entropy by an additive constant: $\bar S_A=S_A+S_0$. The extra entropy $S_0$ could be related to  quantum corrections of the Einstein action, or else to  scale  (or  conformal) invariant  terms added to it in order to describe the degrees of freedom prescribed to the conformal crossover. 
 Gauss-Bonnet (GB) Gravity, with action%
 \footnote{Here $\la$ is a new dimensionless ``gravitational coupling'' constant, while $L$ is a new length scale which can be chosen as  $L = l_{pl}$ by an appropriate redefinition of $\la$; the cosmological constant $\rho_{\Lambda}=\Lambda>0$ is included   in the potential $V(\s)$ of the matter Lagrangian ${\cal L}_{\rm{matter}}$.} 
\br
 S_\GB = \int\! \frac{\sqrt{-g} \, d^4x}{16 \pi G} \Big\{ R - \la L^2 \left[ 2 R^2 - 8 R_{\a\b} R^{\a\b} + 2 R_{\a\b\m\n}R^{\a\b\m\n} \right]
+ {\cal L}_{\rm{matter}} \Big\},\label{action non cf}
\er
represents a quite reasonable example for such an improvement of Einstein Gravity that provides an explicit form of the desirable new constant term (for $\lambda<0$) in the apparent horizon entropy \cite{caikim,caimu,sinha-cosmo,lovecosmo}: 
\begin{eqnarray}
\bar S_A = 8 \pi^2 \, \frac{1}{ H^{2}} \left( 1 - 2 \la L^2 \,  H^2   \right) = S_A + S_0,  \quad\quad S_0=8\pi^2|\lambda|\frac{L^2}{l^2_{pl}}. \label{s(H)} 
\end{eqnarray}  
An important feature of the GB term is that it is \emph{scale invariant}, and being also a \emph{topological invariant} it does not contribute to the equations of motion. Thus the Friedmann Eqs.(\ref{frw}) keep their form and, as a consequence, their SFD properties remain unchanged. Therefore, all the SFD constructions and the CCC and TD restrictions on the matter content we have derived (for Einstein Gravity) in Sects.\ref{Sect.scalarsfd}, \ref{Sect.CCC} and \ref{Sect.TDSFD} are still valid for the GB gravity as well.

With this ``GB renormalization", the total entropy becomes 
\br
& S_t = S_f + S_A + S_0.	\label{totalentropywithGB}
\er
The addition of a constant GB entropy $S_0$ slightly modifies the transformation law (\ref{totepartnonGB}). If we do not impose any additional requirements relating $\ha S_0$ and $\ch S_0$, 
the SFD transformation for the GB improved total entropy (\ref{totalentropywithGB}) takes the from
\br
& \ch S_t = \left(\frac{\ha\rho_{\Lambda}}{\ch\rho_{\Lambda}}\right)^{\frac{3}{4}} \ha S_f + \left(\frac{\ha\rho_{\Lambda}}{\ch\rho_{\Lambda}}\right) \ha S_A \left(\frac{\ha S^{\delta}_{\La}}{\ha S^{\delta}_A} - 1 \right)^{\tfrac{1}{\delta}}+\ch S_0 \label{totepart}
\er
or, in the self-dual case,
$\ch S_t=\ha S_t + \ha S_A \Big [\left(\frac{\ha S^{\delta}_{\La}}{\ha S^{\delta}_A} - 1\right)^{\tfrac{1}{\delta}} -1\Big] +\ch S_0 - \ha S_0$.
The main importance of $S_0$ resides in that we may regard it as an \emph{entropy of the crossover}, related to the specific conformal field theory valid at the immediate vicinity of $\cal X$.
On the ``past'' side of $\cal X$, the pre-big-bang entropy is given by $\ha S_t = \ha S_\La$ (cf. Eq.(\ref{horts})). If there was no GB entropy, on the ``future'' side of $\cal X$, we would have (the problematic) $\ch S_t = 0$. Thus the conformal theory on the crossover would be responsible for a ``complete loss of memory'' between two consecutive aeons. 
The effect of the presence of $\ch S_0$ is, therefore, that it allows us to keep a ``partial persistency of the memory"  across the crossovers.

Let us look in detail at what happens to entropy when one passes between consecutive aeons in a cyclic sequence, say $\cal A_0$, $\cal A_1$ and $\cal A_2=\cal A_0$. We further denote $\cal A_0 = \ha {\cal A}$ and $\cal A_1 = \ch{\cal A}$. 
There are two crossovers $\cal X_0$ and $\cal X_1$, cf. Fig.\ref{RecicledCosmologyAll}. Each has a corresponding GB entropy and we shall assume that the GB scales $\ha L$ and $\ch L$ are \emph{different}, with $\ha L \leq \ch L$ (such that $\ha S_0\leq \ch S_0$), while the GB coupling $\la$ remains unchanged. 
 The total entropies   behave, near the first crossover $\cal X_0$, as $\ha S_t|_{\ha a\to \infty}=\ha S_0+\ha S_{\La}$ and $\ch S_t|_{\ch a\to 0}=\ch S_0$. 
Therefore, the entropy \emph{losses} on each crossover are given by:
\br
&&\Delta S_{\cal X_0} = (\ha S_0+\ha S_{\La}) - \ch S_0 = \frac{4\pi^2 }{l^2_{pl}}\left[2|\la|( \ha L^2 - \ch L^2)+ \ha L^2_{dS}\right] ,\nonumber\\\
&&\Delta S_{\cal X_1} = (\ch S_0+\ch S_{\La}) - \ha S_0 = \frac{4\pi^2 }{l^2_{pl}}\left[2|\la|( \ch L^2 - \ha L^2)+ \ch L^2_{dS}\right] ,\label{jump}
\er
where we have used the de Sitter radii $L_{dS}$, with $\ch \rho_{\Lambda}=3/\ch L^2_{dS}$  and $\ha \rho_{\Lambda}=3/\ha L^2_{dS}$. 
Note that there is one special choice for the GB coupling,
\be
|\la| =\frac{\ha L^2_{dS}}{2(\ch L^2 - \ha L^2)} ,
\ee
that ensures the continuity of the entropy on  $\cal X_0$
(and  on  all the equivalent crossovers $\cal X_0 = \cal X_2 = \cal X_4 = \cdots$), viz. $\Delta S_{\cal X_0}= 0$. But it is impossible to make it continuous on the next (or on the previous) crossover as well,  since now we have  $\Delta S_{\cal X_1}= \tfrac{4\pi^2 }{l^2_{pl}}\left(\ch  L^2_{dS} + \ha L^2_{dS}\right)$. 
In the \emph{self-dual} case, i.e. for  $\ch \rho_{\Lambda}=\ha \rho_{\Lambda}$ and for equal GB scales $\ch L=\ha L$, the losses of entropy  at the different crossovers are identical: $\Delta S_{\cal X_0} = \Delta S_{\cal X_1}= \cdots =S_{\La}$, as well as  its evolution inside all the aeons --- it always increases from $S_0$ to $S_0+S_{\La}$.

An important ingredient defining the  described entropy evolution is the specific choice of GB entropy contributions at different aeons.  Within the framework of the SFD symmetric cosmologies, one can also  consider  another  \emph{semi-infinite SFD extension}, representing a chain of aeons $\cal A_j$ with  \emph{different} GB scales 
$L^2_j=(j +1) l^2_{pl}$ at different aeons. 
This chain begins with a  ``primordial crossover" $\cal X_0$ of minimal entropy $S_0=4\pi^2|\la|$,  and  the total entropy of a future aeon $\cal A_j$ is given by 
\be 
S_{t}(j)=S_f(j)+S_{\La}(j) + (j+1)S_0 , \label{cyctotentr}
\ee
where $S_{0}(j) = (j +1)S_0$ is the GB contribution. The effect of this particular form of the GB entropies $S_{0}(j)$ may be also interpreted as an extension of the SFD entropy definition, that permits an accumulation of the  ``primordial GB entropy" $S_0=4\pi^2|\la|$ along the consecutive self-dual aeons.
We may depict the evolution as follows:
\be
\cal X_0 \;\;  ||\;\;  S_0 \;\;  \hookrightarrow  \;\; S_0+S_{\La} \;\; ||\cal X_1|| \;\; 2 S_0 \;\;  \hookrightarrow  \;\;2S_0+S_{\La} \;\; ||\cal X_2||\;\;  3S_0 \;\;  \hookrightarrow  \;\; 3S_0+S_{\La}\;\;||\cal X_3|| \cdots \nonumber
\ee
It is clear that  all the entropy losses $\Delta S_{\cal X_j}$ are now identical  by construction,
\be
\Delta S_{\cal X_j}= S_{\La}- S_0= \frac{4\pi^2 }{l^2_{pl}}\left(L^2_{dS}- 2|\la|l^2_{pl}\right),\label{cyclicjump}
\ee
and  we have an  \emph{equal partial loss}  of entropy at each crossover. Therefore, we may reach the \emph{continuity} of the  entropy on all  the crossovers  by choosing $|\la| = L^2_{dS}/2l^2_{pl}$. Such a choice ensures not only the continuity, but also the ``eternal" increasing of the entropy (say for $\delta=3/4$ model (\ref{chaply})) considered for the entire infinite chain of aeons.

The above two examples of different choices of the values of the GB parameters   demonstrate how the additional requirement on the form of the GB term at different aeons can give rise to qualitatively different behaviors of the entropy evolution in the SFD symmetric expansion/expansion models. Hence a more complete investigation of the crossover features is needed in order to select one among the many allowed entropy evolutions of such CCC-like models.

\section{Concluding remarks}\label{Sect.conclude}

Veneziano's idea \cite{venez} of using scale factor inversion as a symmetry principle for  constructing   pre-big-bang extensions of the FRW solutions of Dilaton Gravity has been shown to have a successful implementation for pre-big-bang and cyclic cosmologies also in Einstein Gravity (and its Gauss-Bonnet extension), when the  \emph{conformal time SFD} transformations \cite{camara}  are imposed  as a symmetry of these cosmological models.
The possibility of combining the scale factor inversions  with conformal time translations (\ref{ContraExpanDuali}) or with reflections (\ref{ExpanExpanDuali})  gives rise to two distinct  pre-big-bang evolutions --- the contraction/expansion  and the expansion/expansion   ones.

In the present paper we have chosen to study in more detail the geometric and thermodynamical consistency of   SFD symmetric  models of the \emph{expansion/expansion type} and their application in the construction of relevant examples of Penrose's Conformal Cyclic Cosmologies \cite{penrose}. The fact that the employment of the scale factor duality ideas to CCC models is an unexplored area that deserves to be investigated is only one of our motivations. The results presented in Sects.\ref{Sect.CCC} and \ref{Sect.TDSFD} demonstrate the power of the SFD requirement --- when applied to CCC, it allows us to describe all the physical quantities characterizing  the present aeon  in terms of the ones of the past aeon and vice-versa. It also imposes certain restrictions on the evolution of the total entropy of the observable universe in the SFD dual aeons (and on  its  behavior  on the crossover), compatible with the generalized TD second law. Finally, the SFD symmetry  picks out a family of relevant examples of self-dual cosmologies, whose  matter content  coincides with the $SU(1,1)/U(1)$  gauged K\"ahler sigma models with the specific self-interaction (\ref{kahler}). 

Our concluding remarks concern some of the  virtues and the drawbacks  of these pre-big-bang and cyclic  cosmological models, in search for convincing arguments about the relevance of  conformal time SFD as an asymptotic UV/IR symmetry  of the universe evolution. The  comparison of the main features of the  CCC-like expansion/expansion models with their  contraction/expansion relatives aims to clarify the benefits of the employment of the former as unconventional cosmological models. Within the description of the physically consistent realizations of the  restrictions imposed by  scale factor duality,  we  will briefly  discuss some consequences   of the SFD requirement on the properties of  the \emph{adiabatic fluctuations} around the SFD symmetric  FRW backgrounds.

\subsection{More on Contraction/Expansion SFD Models }

As we have shown in  Sects.\ref{Sect.SFDasSymm} and \ref{Sect.scalarsfd}, for each choice  of dual  (or self-dual)  pairs of matter fluids, one finds two kinds of SFD symmetric FRW solutions, corresponding to the two realizations (\ref{ExpanExpanDuali}) and (\ref{ContraExpanDuali}) of the conformal time  scale factor duality  (\ref{sfd}). The different nature of  the pre-big-bang phases of the expanding/expanding (E/E) and contracting/expanding (C/E) solutions lead to very different transitions between  the corresponding dual phases: (i) the big-crunch/big-bang singularity at $\ha a=0=\ch a$,  versus  (ii) the conformal crossover, where one identifies $\ch a=\infty$ with $\ha a=0$. Let us  recall the action of the SFD transformations  used to ``glue"  such pairs  of dual solutions: in  the case of  C/E models, they map the  \emph{initial}  pre-big-bang accelerated contraction  phase  into the  \emph{initial} decelerated expansion phase of the next post-big-bang epoch,  while  for the  E/E models, the \emph{initial} decelerated  expansion phase of the past aeon is the SFD ``image"  of the \emph{final} accelerated expanding phase in the future aeon and vice-versa.

 The SFD symmetric C/E models, studied in Sections \ref{Sect.SFDasSymm} and \ref{Sect.scalarsfd},  provide examples of contraction/expansion FRW backgrounds with a reasonable matter content satisfying the weak energy condition.
The \emph{self-duality} of the matter potential $V(\s)$ (\ref{poten}), defining  cosmological models with modified Chaplygin gas EoS (\ref{chaply}),  is the main responsible for the continuity of the scale factor $a(\eta)$, of the scalar field $\s(\eta)$ and of their derivatives at the big-crunch/big-bang singularity at $\eta=0$. Their  \emph{partially self-dual} counterparts  offer an interesting C/E cosmological scenario with initial  and final ``de Sitter states"  of different cosmological constants $\ch\rho_{\Lambda}\neq\ha\rho_{\Lambda}$. However, the derivatives of the scale factors at the singularity now present finite jumps  due to the different  relative radiation densities $\ch\rho_r\neq\ha\rho_r$. 
It is worthwhile to mention that these features are also shared by  the C/E cosmology involving  $\Lambda$CDM model and its SFD dual  (\ref{LaCDMdaul}).  

The cosmological feasibility  of the  contraction/expansion models depends  on the   eventual instabilities and ghosts problems  that might be caused  by  the requirement of scale factor duality   as an  \emph{asymptotic  symmetry} of  the universe evolution.  The  appearance of such inconsistencies (and the ways they might be avoided) are mainly related to the properties of the adiabatic fluctuations around the \emph{contraction phase} of the SFD symmetric backgrounds.  The similarity of certain features of  our C/E models with those  of  the Pre-big-bang Scenario, and the Ekpyrotic and Cyclic cosmologies, suggests that we can make the discussion of their flaws shorter by borrowing  and readapting some of  the arguments  from the extensive critical analysis of those models presented in Refs.\cite{brandenberger2016bouncing, linde-cyclic-rev, venez2, steinhardt2002cosmic, kallosh2008new, hsu2004gradient}.  One should stand out a few of the problems  to be faced 
--- the chaotic anisotropic Belinskii-Khalatnikov-Lifshitz-like (BKL) singularities \cite{belinskii1970oscillatory,damour2003cosmological}; the gradient and thermodynamical instabilities and  the absence of ghosts, phantoms and other WEC breaking inconsistencies.

The cosmological models whose EoS parameter is bounded as $w < 1$ in the  \emph{contraction phase} are known to develop chaotic  BKL ``mixmaster'' behavior near the singularity.  Therefore all the considered  SFD dual and self-dual C/E models indeed  suffer from BKL inconsistency,%
\footnote{We have to remind  that the SFD models  with EoS (\ref{chaply}) for $\delta<0$ are in fact non-singular but represent certain TD consistency problems \cite{sotkov2016thermodynamics} and due to $v_s^2<0$ they might develop  gradient instabilities as well.}
due to   the ``SFD restrictions''  $-1\leq w(\rho)\leq 1/3$ imposed on the  weak energy conditions.

In the  ekpyrotic/cyclic  models \cite{steinhardt2002cosmic} the BKL chaotic singularity is avoided via a mechanism that makes the (effective) EoS parameter $w \gg 1$ as the universe collapses. A new   \emph{ekpyrotic} (with low density and low curvature) slow contraction  phase is created by a specific  scalar field potential with a negative valley, or else  by  considering a  ``ghost condensate"  matter content  \cite{lehnert,new-ekp}. Although they are not free from certain ``ekpyrotic  ghost" inconsistencies \cite{kallosh2008new}, they turn out  to provide a reasonable contracting pre-big-bang alternative of the standard inflation scenario.
It is then natural to try to incorporate  the particular features of  these models  within SFD symmetric C/E cosmologies, in order to make them  free of BKL singularities. However, the SFD dual of an ekpyrotic phase with $\ha w\gg 1$ must have $\ch w = - \ha w - \tfrac{2}{3} \ll -1$. So it turns out that the WEC is then strongly violated in the expanding post-big-bang phase of these ekpyrotic SFD models and, as a consequence, they are subject to dangerous ``phantom''  \cite{carroll2003can, kallosh2008new} and  gradient  instabilities  \cite{hsu2004gradient}. We have to also notice that the SFD symmetric ekpyrotic cosmologies  \emph{do not} reproduce one of the basic features of the original ekpyrotic/cyclic models, namely that their post-big-bang phase should be given by a standard $\Lambda$CDM model. 

In brief,  the  problems  of  SFD symmetric  C/E models  demonstrates that their  ekpyrotic (BKL free) improvements  also \emph{fail}   to represent consistent cosmologies, due to the presence of ghosts and gradient instabilities induced by the SFD requirement.

\subsection{On the scale factor duality of  adiabatic fluctuations}

Fluctuations around   the  homogeneous and isotropic backgrounds are expected to break  down the scale factor duality, mainly due to the extra scales involved in the observed non-homogeneous and anisotropic evolution of the universe. In the case of  SFD symmetric C/E models, the construction of a  smooth bounce  (that should replace the singularity)  also necessarily introduces certain higher derivative ``quantum corrections" to Einstein gravity (i.e. powers of the curvature invariants), that are  known to be  \emph{incompatible} with the scale factor duality.
  Such a problem is, however, \emph{absent} in the considered  CCC-like models, due to the special features of the \emph{conformal crossover} as the transition  between two consecutive aeons. The cosmological applications of such models and the use  of their late time accelerated phase as a ``pre-big-bang inflationary" phase  crucially  depends on the existence of  a consistent mechanism for the  transmission of  the fluctuations through the conformal crossover.

The well known S-duality extension  \cite{brustein1998duality} of the original SFD symmetry%
\footnote{See sect.2 of ref.\cite{camara} for a comparison of its  similarities and differences with the considered conformal time SFD (\ref{sfd}).}
 of  Dilaton Gravity provides an indication about the possible realizations of the SFD transformations of the tensor modes (i.e. of helicities $\pm2$) of metric fluctuations $g_{\pm}(\eta,x_i)\approx a^2(\eta)(1+ h_{\pm}(\eta,x_i))$ that keep invariant their energy density. By considering the second order terms in the expansion of the Einstein Gravity action, one can derive the contribution of the tensor fluctuations to the corresponding 
  Hamiltonian density:
\br
\cal H = \frac{1}{2} \int \! d^3 x \left[ \frac{\Pi^2}{a^2(\eta)} + a^2(\eta) \left( \nabla h \right)^2 \right] = \frac{1}{2} \int \! d^3 k \left[ \frac{|\Pi_k|^2}{a^2(\eta)} + k^2a^2(\eta)  |h_k| ^2 \right], 
\er
where $h$ is either one of the  transverse traceless modes $h_{\pm}$ (i.e. $h_j^j=0$ and $ \nabla^jh_{ji}=0$) of the gravitational wave; $\Pi= a^2h' $  their canonically conjugate momenta; $h_k$, $\Pi_k$  the corresponding Fourier modes (with wavenumber $k$)  and the 
abreviations  $|h_k| ^2=h_kh_{-k}$ and $k^2=|k|^2=k_i k_i$ have been assumed.
 It may be easily verified that the Hamiltonian and, consequently, the equations of motion for the fluctuations $h_k$ and for  their momenta $\Pi_k$   are invariant under the transformations
\br
a \mapsto \tilde a(\til \eta) = c_0^2 / a(\eta) \; , \quad h_k \mapsto \tilde h_k(\til\eta) = - \tfrac{1}{c^2_0 \, k} \Pi_k(\eta) \; , \quad \Pi_k \mapsto \tilde \Pi_k(\til \eta) = c^2_0 \, k h_k (\eta),		\label{SFDtensorfluct}
\er
consisting of the scale factor inversion (\ref{sfinv}) along with an appropriate exchange of the fields  with their momenta. 
Such  an enhancement  of the SFD symmetry by a particular  canonical transformation allows to completely determine the tensor modes $\ch h_k(\eta)$ of  the metric fluctuations in the early time (radiation/matter dominated)  post-big-bang phase in terms of their SFD dual (late-time) accelerated pre-big-bang counterparts  $\ha h_k(-\eta)$, 
\br
\ch h_k(\eta) =  \frac{1}{c^2_0 \, k} \ha a^2(-\eta) \ha h'_k(-\eta), \quad\quad\quad 0\leq\eta\leq\eta_f.\label{tensorsym}
\er
It remains however to establish  the prescription of how to follow these metric perturbations through the crossover surface $\cal X$, i.e. to  define appropriate matching rules for the solutions $\ch h_k$ and $\ch h_k$ at $\eta=0$.
The SFD symmetry requirement (\ref{tensorsym}), together with the natural condition for (asymptotic) vanishing of the fluctuations $\bar h(\eta)= a(\eta)h_k(\eta )$ at the homogeneous and isotropic initial/final stages of the universe evolution, can be satisfied by imposing $\ha{\bar h}_k(0)=0=\ch{\bar h}_k(0)$ and permitting a finite jump $\Delta\bar h_k=\ha{\bar h}'_k(0) - \ch{\bar h}'_k(0)$ of their derivatives. It is worthwhile to mention that in the  case of  $\delta=1/2$ self-dual model, where  the fluctuation equations are exactly solvable, the above ``crossover conditions"  allow  to  determine the  ($k$-dependent) coefficients  of the solutions for $\ha h_k(-\eta)$, $\ch h_k(\eta)$ and  the exact form $k^2_n$ of their \emph{discrete} spectrum as well.

The  scalar modes  $\Psi(\eta,x_i)$ of the metric perturbations in longitudinal gauge (and in the absence of anisotropic stresses), $ ds ^2= a ^2(\eta) [ - (1 +  2\Psi) d\eta^2 + (1-2\Psi)\delta_{ij}dx^i dx^j ]$
obey the equations 
\be
 u''_k +[k^2- \frac{z''}{z}]u_k=0,\quad\quad u_k=z \Psi_k,\quad\quad z=a^2 \frac{\sigma'}{a'},\label{scmode}
\ee
that are similar to the ones for the tensor modes, when written for $\bar h_k= a h_k$. However the difference between their  Barden potentials ($V_{\bar h}=\tfrac{a''}{a}$  and $V_u= \tfrac{z''}{z}$)   as well as  the  different SFD transformation properties of the variables  $a$ and $z$ 
\emph{do not permit}\footnote{With an exception of the particular cases, when $z\sim a$, as for example for perfect fluids.} to extend  the \emph{background scale factor duality} (\ref{sfd})  as we have done for the corresponding tensor modes. 
Although now one cannot  use the SFD symmetry in order to define the post-big-bang scalar fluctuations in terms of the corresponding pre-big-bang ones,  an important question to be addressed concerns the  consequences  of their SFD dual matter content (and  of  the specific SFD symmetric  backgrounds) on  the properties of  fluctuations and  of their spectra. 
The  \emph{background SFD symmetry} in fact completely determine  the potentials $\ha V_{\ha u}(\ha z)\neq \ch V_{\ch u}(\ch z)$ of the scalar mode fluctuation equations at two consecutive aeons. For example,  the near crossover approximation  of the potentials $V_u= \tfrac{z''}{z}$  corresponding to the self-dual SFD models (\ref{chaply})  takes a rather familiar (but different)  form at the both sides of the crossover surface $\cal X = [\{ \ha a \to \infty \} \sim \{ \ch a \to 0 \}]$:
\br
\ha u''_k +[k^2 - \frac{\nu^2-\frac{1}{4}}{\eta^2}] \ha u_k=0,\quad\text {for} \quad \ha\eta \rightarrow 0^-,\quad\text{and}\quad \ch u''_k + k^2 \ch u_k=0,\quad\text {for} \quad \ch \eta \rightarrow 0^+ \label{scmodesapprox}
\er
(with $\nu=2\delta-\tfrac{3}{2}$), which are derived  by using  a simplified  $z=-2 \left(\tfrac{\rho_r}{3}\right)^{\tfrac{\delta}{2}} a^{1-\delta} (aH)^{-\delta}$ form  in the evaluation of the limits  $\ha a\rightarrow \tfrac{1}{\eta}$ and $\ch a\rightarrow \eta$.
Their solutions are indeed well known and it seems quite reasonable  to impose for   $\ha u_k$ and $\ch u_k$  a  crossover prescription similar to the one used  for the tensor modes.
Although such a choice provides  consistent answers for   $\delta=1$ and $\delta=1/2$ self-dual models,  the derivation of the  generic form of the scalar modes crossover matching conditions requires  further investigations.

The proper structure of all the SFD symmetric CCC models  opens a room for certain speculations about the  eventual double role of the  accelerated asymptotically $dS_4$ phase at the vicinity of crossover.  One may try to derive the restrictions  under which it can serve as a  pre-big-bang inflationary phase  of superluminal expansion and cooling up to almost zero temperature, and even more ---  that the conformal crossover could play the role of a kind of reheating transition to the radiation dominated (high temperature) post-big-bang  phase. The solutions of the horizon and flatness problems  in these considered CCC models are quite similar to those offered by  the well known dilaton gravity pre-big-bang models and some of the cyclic models as well. The question concerning the  existence (or not) of conditions favorable  for the complete realization of such pre-big-bang inflationary scenario  is, however,  out of the scope of the present paper.

\subsection{Towards the conformal crossover holography}

As we have shown in Sect.\ref{Subsect.Crossover}, the origin of   the asymptotic SFD symmetry of CCC models is closely related to  the ``reciprocal hypothesis" and to the  \emph{conformal crossover}, i.e. the existence of conformal (Weyl) equivalence class of metrics defining the transition between the future infinity of one aeon to the big-bang singularity of the next one.  We  have deduced  a set of SFD  requirements  concerning the \emph{conformal properties} of the  crossover and also  a particular mechanism for  the explicit  breaking of this crossover's conformal symmetry, namely  by  adding  certain relevant operators  preserving unbroken the subgroup of scale factor inversions. We  consider these results  as a step towards the realization  of   the   4D conformal field  theories%
\footnote{Which might contain Weyl gravity or/and  the conformal coupling of a scalar field to Einstein gravity, with conformal invariant self-interaction.} describing  these transition  regions. The similarity of the primordial conformal crossover, introduced in Sect.\ref{Sybsect.GBcross}, to the original ideas of conformal and pseudo-conformal initial stage of  the universe evolution \cite{khoury1,motolla, antoniadis2012conformal, libanov2015towards,kaloper}, suggests that  they could be realized within the frameworks of the SFD symmetric CCC models as well.

Another conformal byproduct  of our investigation of  the SFD/CCC correspondence  are the following two  \emph{indications} of how one can implement  the  $dS_4/CFT_3$  holographic methods%
\footnote{And their off-critical  (perturbed $CFT_3$)/(asymptotic $dS_4)$ version \cite{van-der, van2004inflationary,skender-cosmo}} \cite{stromi2, strominger2001inflation}  for the description of  the conformal  crossovers in terms of  certain \emph{euclidean  three dimensional conformal theories}. The first one is the specific holographic relation  
$$ m^2_{\s} L^2_{dS}= 2\delta (3 - 2\delta)$$ 
between the 4D de Sitter scale $L^2_{dS}$  and 4D scalar field mass $m_{\s}$  with the anomalous scaling dimensions $\Delta_{\s}=2\delta$  of the  perturbing relevant operators with $ 2\delta<3$ in the corresponding $CFT_3$. The second one concerns  the  value of the central charges
$${\bf{a}}_{E} = S_{\Lambda} =4\pi^2 \frac{L^2_{dS}}{l^2_{pl}}={\bf{c}}_E$$ 
of  these crossovers euclidean $CFT_3$'s,  directly related to the asymptotic limit of the corresponding horizons entropy  \cite{sinha-cosmo} or its GB improved version \cite{caimu,sinha-cosmo,lovecosmo}:
\be
{\bf{a}}_{GB}(j)=S_{0j} +S_{\Lambda}=\frac{4\pi^2 L^2_{dS}}{l^2_{pl}}\left(1+ \frac{2|\la| L^2_j}{L^2_{dS}} \right),\label{centralcharge}
\ee
 but now with ${\bf{a}}_{GB}\neq{\bf{c}}_{GB}$ (see \cite{sinha-cosmo,skender-cosmo}). According to our discussion in Sect.\ref{Sybsect.GBcross} above, the values of the central charges ${\bf{a}}_{GB}(j)$  of the sequence of $CFT_3(\cal X_j)$ are determined by the specific choice of the GB scales $L_j$ at different aeons $\cal A_j$, related to the conditions imposed  on the \emph{entropy jumps} or by requiring its continuity on each crossover (when it is possible). For example, in the case of equal GB scales $L_j=L\propto l_{pl}$ all the crossovers are represented by a unique  $CFT_3$,  which in the CCC framework corresponds to the complete loss of  the entropy $S_{\Lambda}$ created at the precedent cycle. No primordial crossover needs to be introduced in this case, and indeed the  evolution of such SFD symmetric CCC model  turns out to be  completely cyclic.

 We should mention in conclusion  that   our partially self-dual CCC-like models   also provide  all the remaining  ingredients  needed for the complete realization of their holographic description in terms of the ``RG flows" in the dual perturbed $CFT_3$  \cite{van-der, van2004inflationary,skender-cosmo}}, namely  --  their  exact beta functions,  the explicit form of the corresponding relevant (and irrelevant) 3-d operators  
 and finally  -  the non-perturbative form of their correlation functions  that can be used for the holographic reconstruction  of   the  density fluctuations spectrum of such CCC-like cosmological models.

 \vspace{0.3cm}
 
\textbf{Acknowledgments.} ALAL  thanks CAPES (Brazil) for financial support. The research of UCdS is partially supported by FAPES(Brazil).
 
 \appendix

\section{Cyclic  SFD models}\label{AppA}

Here we give an explicit example of the construction of a chain of dual aeons described in Sect.\ref{Sect.SFDasSymm}, representing a cyclic evolution of  an universe with deceleration and acceleration periods and with a finite conformal-time lifespan. A specially appropriate case is the Chaplygin gas (\ref{chaply}), with $\delta = 1/2$. As stated in Sect.\ref{Sect.CCC}, it satisfies the SRM hypothesis. Also, it is partially self-dual, as described in Sect.\ref{SubSectDualPatDual}. 
The scale factor evolution may be readily integrated:
\br
 a_1(\eta_1) =  \left( \tfrac{\rho_{r1}}{\rho_{\Lambda1}}  \right)^{\frac{1}{4}} \tan \left[ \tfrac{\pi}{2} \eta_1 / \eta_f \right] . \label{a1exApp}
\er
where
$\eta_f = \frac{3\pi}{2 (\rho_{r1} \rho_{\Lambda1})^{1/4}}$.
This aeon, $\cal A_1$, has a big-bang at $\eta = 0$ and ends at $\eta = \eta_f$, so $\eta_1 \in [0,\eta_f]$.  We may use Eq.(\ref{SFIrecurs}) to find an adjacent aeon, either $\cal A_0$ or $\cal A_2$. Let us choose the former, with $a_0(\eta_0) = c_0^2 / a_1(-\eta_0)$, i.e.
\br
a_0(\eta_0) =  c_0^2 \left( \tfrac{\rho_{\Lambda1}}{\rho_{r1}} \right)^{\frac{1}{4}} \cot \left[ - \tfrac{\pi}{2} \eta_0 / \eta_f \right] = \left( \tfrac{\rho_{r0}}{\rho_{\Lambda0}} \right)^{\frac{1}{4}} \tan \left[ \tfrac{\pi}{2}(\eta_0 + \eta_f) / \eta_f  \right]. 
\label{a0exapp}
\er
In the last equality, we have used a trigonometric identity, $\cot(-x) = \tan(x+\pi/2)$, and the relation (\ref{rhochanges}) between the parameters. Comparison between this equality and (\ref{a1exApp}) illustrates explicitly the partial self-duality of the model: the dual functions $a_1(\eta_1)$ and $a_0(\eta_1)$ have exactly the same form, the difference being only in the \emph{values} of the parameters $\{\rho_{r1}, \rho_{\La1}\}$ and $\{\rho_{r0}, \rho_{\La0}\}$.


\begin{figure}[ht] 

\centering
\includegraphics[scale=0.55]{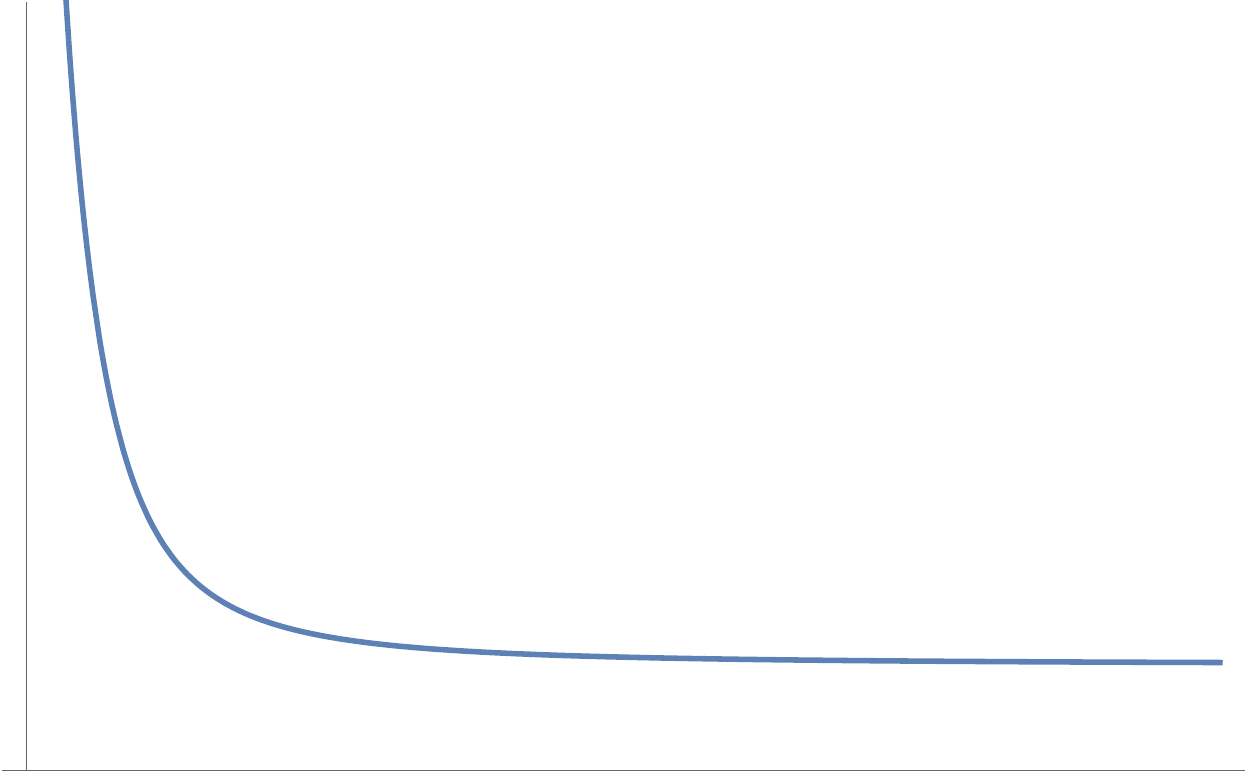} \label{delta_3_4}
\caption{
Profile of the function $f(a)$ for $\delta = 3/4$. 
}\label{Sproof}
\end{figure}


It is instructive to find now $\cal A_{-1}$, since according to the $Z_2$ property of (\ref{SFIrecurs}) its scale factor must be identical with (\ref{a1exApp}), including the same values of the parameters. Indeed, we have
\br
a_{-1}(\eta_{-1}) = \frac{c_0^2}{a_0(-\eta_{-1} - 2 \eta_f)} = \left( \tfrac{\rho_{r1}}{\rho_{\Lambda1}} \right)^{\frac{1}{4}} \tan \left[ - \frac{\pi}{2} \frac{ (\eta_{-1} + 2 \eta_f) }{\eta_f} \right]  = \left( \tfrac{\rho_{r1}}{\rho_{\Lambda1}} \right) ^{\frac{1}{4}}\tan \left[ \tfrac{\pi}{2} \eta_{-1} / \eta_f \right] \nonumber
\er
as expected. In the last equation we have used  a property of the periodicity of the $\tan x$ function to demonstrate our point that $a_{-1}(\eta_{-1}) = a_1(\eta_1)$.

The explicit use of trigonometric identities and the special property of the model being partially self-dual may be misleading, suggesting that our ``chain of aeons'' could only be constructed for such periodic solutions for the scale factor. This is, of course, not the case. To illustrate that it is not so, consider now a generic scale factor
$a_1(\eta_1) = g(\eta_1)$,
where $g$ is an appropriate function, i.e. $g(0) = 0$ e $g(\eta_f) = \infty$ for some $\eta_f$. We have
$$a_2(\eta_2) = \frac{c_0^2}{g( - \eta_2 +2 \eta_f)} ,$$
and
$$ a_3(\eta_3) = \frac{c_0^2}{a_2(-\eta_3 + 4 \eta_f)} = g(\eta_3 - 2\eta_f) ,$$
 which is the original function with a shifted argument. The important point here is that the domain of $g(x)$ is always the \emph{same} interval of the real line, namely $x \in [0, \eta_f]$, although the domain of $a_1(\eta_1)$, $a_2(\eta_2)$,  etc. are indeed different from each other. Therefore the ``periodicity'' of the chain has nothing to do with an eventual periodicity of $g$.

\section{Proof of the $2^\nd$ law for modified Chaplygin gas models}\label{Sect.delta}

The $2^\nd$ law for the generalized entropy inside the apparent horizon in a universe filled with a modified Chaplygin gas (\ref{chaply}) is guaranteed within the range of validity of  the inequality (\ref{col}) only. Clearly it is equivalent to the positivity of the term in the brackets. Let us denote this term by
\br
f(a) = 1+ \frac{\sqrt{3} \mathcal{S}}{8\pi \rho^{\delta}_r} \;   \frac{\left(\rho^{\delta}_r - \rho^{\delta}_{\Lambda}  a^{4\delta} \right)}{a^3 \,\left(\rho^{\delta}_{\Lambda}+\rho^{\delta}_r \; a^{-4\delta}\right)^{\frac{1}{2\delta}}} . 	\label{f}
\er
We will be concerned only with the physically reasonable range of $\delta$, considered in Sect.\ref{Sect.TDSFD}, viz. $\frac{3}{4} \leq \delta \leq 1$. We must have $f(a) > 0$ for all $a > 0$. We first take the limit $a \to \infty$ and realize that  $f > 0$ is equivalent to $a^{4\delta - 3 } <  \frac{8\pi \rho^{\delta}_r}{\sqrt{3} \mathcal{S} \rho_\La^\delta }$. This inequality is however violated for sufficiently large values of $a$ whenever $\delta > \frac{3}{4}$. 
Therefore the only possibility that remains is $\delta = \frac{3}{4}$. In this case, it is immediate to verify that $f(a)$ is a monotonic decreasing function, cf. Fig.\ref{Sproof}. The asymptotic inequality 
$$1 <  \frac{8\pi \rho^{3/4}_r}{\sqrt{3} \mathcal{S} \rho_\La^{3/4} },$$
(again for $a \to \infty$) becomes identical to the stated one (\ref{3/4}), once one takes into account the Steffan-Boltzmann law to write $\cal S = \tfrac{4}{3} (4 \s)^{1/4} \rho_r^{3 / 4}$ (cf.  Eq.(\ref{r_T})).


\bibliographystyle{JHEP}

\bibliography{References}

\end{document}